\documentclass[10pt, a4paper]{article}
\usepackage{jheppub}
\usepackage{amsmath ,amsthm ,amssymb}
\usepackage{graphicx}
\usepackage{relsize}
\usepackage{bbm}
\usepackage{mathtools} 
\usepackage[utf8]{inputenc}
\usepackage{stmaryrd} 
\usepackage{tikz-cd}
\usepackage{tikz}
\usetikzlibrary{matrix} 
\usepackage{bbold}
\usepackage{cleveref}

\usepackage{xcolor}

\theoremstyle{plain}
\newtheorem{thm}{Theorem}[section] 
\newtheorem{prop}[thm]{Proposition}

\theoremstyle{definition}
\newtheorem{defn}[thm]{Definition} 

\title{A gravitational action with stringy $Q$ and $R$ fluxes via deformed differential graded Poisson algebras}
\author[*,**]{Eugenia Boffo,} \author[**]{\, Peter Schupp}
\affiliation[*]{Charles University Prague, Faculty of Mathematics and Physics\\ 
Mathematical Institute, Prague 186 75, Czech Republic}
\affiliation[**]{Jacobs University Bremen, Department of Physics and Earth Sciences\\ Campus Ring 1, 28759 Bremen, Germany}

\emailAdd{e.boffo@jacobs-university.de} 
\emailAdd{p.schupp@jacobs-university.de}
\abstract{
We study a deformation of a $2$-graded Poisson algebra where the functions of the phase space variables are complemented by linear functions of parity odd velocities. The deformation is carried by a $2$-form $B$-field and a bivector $\Pi$, that we consider as gauge fields of the geometric and non-geometric fluxes $H$, $f$, $Q$ and $R$ arising in the context of string theory compactification. The technique used to deform the Poisson brackets is widely known for the point particle interacting with a $U(1)$ gauge field \cite{Tanimura_1992}, {}but not in the case of non-abelian or higher spin fields. {}The construction is closely related to Generalized Geometry: With an element of the algebra that squares to zero, the graded symplectic picture is equivalent to an exact Courant algebroid over the generalized tangent bundle $E \cong TM \oplus T^{*}M$, and to its higher gauge theory. {}A particular idempotent graded canonical transformation is equivalent to the generalized metric.  Focusing on the generalized differential geometry side we construct an action functional with the Ricci tensor of a connection on covectors, encoding the dynamics of a gravitational theory for a contravariant metric tensor and $Q$ and $R$ fluxes. We also extract a connection on vector fields and determine a {non-symmetric {}metric gravity theory} involving a metric and $H$-flux.
}
\begin{document}
\maketitle

\section{Introduction}


An interacting theory for a particle or a field can be formulated starting from the canonical Poisson algebra of functions for the free theory {}by deforming this algebra along force-carrying fields. The technique essentially consists in a {}local redefinition of  phase space coordinates in the direction of gauge and other fields. {}We will apply the method to gravitational interactions and higher gauge fields. The symplectic $2$-form looks quite different in the new chart. Coupling gauge fields to matter field(s) is relatively easy and makes the theory covariant. {} Along these lines, Jackiw \cite{Jackiw:1984rd} first noticed that in a quantum mechanical system unitary representations of the $\mathfrak{u}(1)$ algebra do not associate when a magnetic monopole is present. {}Another pioneering example where this technique is applied is the $N=2$ superparticle with a background~\cite{Dai:2008bh}, where however it was found that for spacetime symmetries the field strengths are trivial and therefore do not propagate. In this article we show how this problem can be overcome in the framework of differential graded Poisson algebras. We will consider the degree-$2$ Poisson algebra of a point particle. Our reference manifold is the (degree-)shifted cotangent space $T^{*}[2]T[1]M$ of the shifted tangent space to a manifold $M$. This is the minimal symplectic realization of a given graded manifold $T[1]M$; for details on how to construct the bundle and relations to other graded manifolds we refer to \cite{1410.3346}. We couple the particle to $g-B$, where~$g$ is a Riemannian metric on the {}base manifold and $B$ a $2$-form, and to $ \left(g+B\right)^{-1} \equiv G^{-1} + \Pi$. The antisymmetric tensors $B$ and $ \Pi$ can be regarded as the gauge fields for some $3$-tensors (with covariant and contravariant indices) arising in string theory compactification with fluxes.
Behind the possibility to pick up and study a closed non-degenerate $2$-form, path-connected to the canonical symplectic form $\omega$, there is Moser lemma \cite{Moser1965}. More precisely, the lemma states that two non-degenerate and closed $2$-forms on a manifold, which can be smoothly deformed into each other, are linked by a diffeomorphism $\varphi_{t}$, for each real $t\in [0,1]$. The forms are cohomologous in $H_{\text{dR}}^{2}$ hence they can only differ by an exact $2$-form $\mathrm{d}A_{t}$. The diffeomorphism is unique up to gauge transformations of $A_{t}$. With {}slight adaption, as we shall see, Moser lemma subsists also when the manifold is graded. $A_{t}$ comprises the gauge fields (but also ``higher'' gauge fields in the geometric ladder), which are therefore incorporated into the Poisson algebra.

Standard exact Courant algebroids are naturally associated with the graded Poisson algebra on the graded symplectic manifold $T^{*}[2]T[1]M$, under the condition that a nilpotent differential operator is also present. Algebroids are the generalization of an algebra, extended over all the points of a manifold; they intrinsically carry the notion of a fiber bundle $E \xrightarrow{\pi} M$. The bracket for the Courant algebroid (CA) is generally taken to be non-skewsymmetric, in order to fulfill Jacobi's identity, and there is also a fiberwise symmetric pairing that respects some compatibility conditions with the bracket. See the article \cite{Gualtieri:2007ng} for a good mathematical explanation of this concept. Furthermore, the bracket accounts for the current algebra of closed bosonic strings \cite{Alekseev:2004np}. When the CA on $E$ is exact, it was shown \cite{Severa:2017oew} to be canonically isomorphic to a CA on the bundle $TM \oplus T^{*}M$. If moreover the metric is the standard pairing of vector fields with $1$-forms, $TM \oplus T^{*}M$ is seen as a principal bundle with structure group $O(d,d)$, that takes into account the stringy T-dualities, which constitute the Weyl group $O(d,d; \mathbb{Z})$. The defining objects (alongside with their relations) of an exact CA on $TM \oplus T^{*}M$ can be obtained from the derived bracket of the Poisson brackets, with the canonically related degree-shifted Hamiltonian vector field Q. The vector field can be used to define a cochain complex, because for the Hamiltonian function $\Theta$, $\text{Q} = \{ \cdot, \Theta\}$, the following master equation holds:
\[
\{ \Theta, \Theta \} = 0.
\]
Introducing additional geometric data on the exact CA we manage to derive a pair of low energy effective actions for string theory compactification, involving  T-dual stringy fluxes.

T-duality is a symmetry exhibited by the non-linear sigma-model of the relativistic bosonic string: In the presence of isometry directions, the sigma-model with background fields $(g,B)$ was shown to be equivalent to the sigma-model based on another pair of backgrounds. This equivalence is dubbed T-duality \cite{Buscher:1987sk}. The famous Buscher rules regulate the correspondence between the fields; in general, T-dual backgrounds do not share the isometry symmetries of the original fields. Moreover, the T-duality symmetry concerns just the effective field theory for the string (it holds at $\left(\alpha'\right)^{0}$ order). In type II string theories, compactification with fluxes on a $\mathbb{T}_{3}$ complex torus was suggested as an alternative to Calabi-Yau compactifications without fluxes, since the additional structure carried by the fluxes can attempt to stabilize the moduli, by making them massive through a potential that combines nicely the superpotential and the K\"{a}hler potential. The fluxes were also believed to shed light on the presence of de Sitter solutions, but research in this direction has not been quite fortunate. In \cite{Shelton:2005cf} it was argued that the NS-NS and R-R fields of IIA and those of IIB must be T-dual to each other, since the two theories are already their respective T-dual when compactified without fluxes. Successive T-duality transformations applied to the Neveu-Schwarz $3$-form $H$ lead to the $f, Q, R$ fluxes according to the chain:
\[
H_{abc} \xrightarrow{\text{T}^{c}} f_{ab}{}^{c} \xrightarrow{\text{T}^{b}} Q_{a}{}^{bc} \xrightarrow{\text{T}^{a}} R^{abc},
\]
where for the last arrow one invokes Buscher rules despite the lack of an isometry direction. This {}results in a rupture with standard geometry, since now the local open sets of the base manifold cannot be patched together; neither with general linear transformations as transition functions, nor with a more general $O(d,d)$ transformation such as a T-duality (like when there is a non-trivial $Q$-flux, which presents this type of monodromy). Despite this unusual feature, the non-geometric flux $R$ is {}very interesting due to its r\^{o}le as source for non-associativity of the phase space algebra of open strings ending on D-branes. {}In more recent work, there are indications that the geometry of Yang-Mills theory on a D3-brane in the background of a $R$-flux can also be understood in a  fully associative picture \cite{Hull:2019iuy}. However conjectures of non-associativity are still viable for closed strings. For a study of non-geometric backgrounds from the string sigma-model perspective a good reference is \cite{Halmagyi:2009te}. Topological T-duality with $H$ flux is explained in \cite{Bouwknegt:2003vb}. Other types like non-abelian T-duality and Poisson-Lie T-duality constitute an active research field. A manifestly T-duality covariant formalism was developed in \cite{Hohm:2010pp} and led to the formulation of the most general invariant action for the background fields, dubbed Double Field Theory (DFT). The Supergravity action for the gravity multiplet $(g,B,\phi)$ is included in the DFT action. From the latter one can also extract an action functional with the $R$-flux as invariant tensor and the $Q$-flux as connection coefficient \cite{Andriot:2012an}. A further equivalent action named $\beta$-supergravity appeared in \cite{Andriot:2013xca}. It was initially obtained from a field redefinition of the NS-NS action functional limit of the DFT covariant action (with $H$-flux and dilaton) but eventually it was expressed through the differential-geometric objects of a Lie algebroid structure on $T^{*}M$ in \cite{Blumenhagen:2013aia}, where $M$ is a regular manifold. Besides Double Field Theory, the fluxes can also be analyzed in the context of Generalized Geometry and algebroids. The bound between T-dual non-geometric fluxes and the Courant bracket is discussed in \cite{Grana:2008yw} which primarily stresses the importance of Generalized Geometry for flux compactification. Reference \cite{Plauschinn:2018wbo} reviews this particular aspect too and provides an exhaustive overview of non-geometric fluxes. In the context of differential graded (dg) symplectic geometry, the fluxes are usually understood as a twist of the differential, see for example \cite{Heller:2016abk}; subsequently at the level of algebroids the bracket gets twisted as well. The fluxes can also arise from an appropriate basis choice, as done for example in~\cite{Blumenhagen:2012nk}. The net result is that the gauge algebra comprises now the fluxes of the T-duality chain as structure functions. The Jacobi identity leads to the Bianchi identity for the fluxes. For a derivation of the string effective action with $H$-flux and dilaton in the context of Generalized Geometry please refer to \cite{Jurco:2015bfs} and \cite{Jurco:2017gii}. 

We are going to adopt the viewpoint of \cite{Blumenhagen:2012nk}, but applied to a deformed $2$-graded Poisson algebra. When switching to the Courant algebroid picture, the Hamiltonian vector field will remain the untwisted differential of the free theory. Although normally in dg-symplectic geometry the standard prescription for the fluxes requires the differential to be twisted by them, the local expression for $H$, $Q$ and $R$ will anyway pop up in our formulas, thanks to other new geometrical data introduced at the level on the generalized tangent bundle in \cref{B}.

The article is organized as follows: we start in section \ref{A} with an overview of the features of the differential graded symplectic manifold $(T^{*}[2]T[1]M, \omega, \text{Q})$. A general deformation is proposed and the outcome is analyzed; the particular example of the deformation that involves $g, B, G^{-1}$ and $\Pi$ is given at the end of this  section. We then proceed in section \ref{CAA} by focusing on the structure of Courant algebroids on the generalized tangent bundle $TM \oplus T^{*}M$, which is equivalent to the graded Poisson structure previously considered. When the latter is deformed, the former gets changed in the same way: We hence unveil the expressions for the new objects in this section. Section \ref{B} is the main body of the article: There we make full use of the developments from the previous sections, in particular the new deformed Dorfman bracket. Defining a generalized Lie bracket alongside with a generalized torsion, naturally leads to an affine connection with torsion. We compute the generalized Riemann tensor for vectors and, separately, for $1$-forms. In this case, the Riemann tensor is a section of $\bigotimes^{3} TM \otimes T^{*}M$, and expresses the curvature of the Riemannian manifold $M$ with metric $G^{-1} \in \vee^{2} TM$. The connection on forms, in light of the ansatz chosen for the deformation, depends on the local expression of the non-geometric $R$ and $Q$, which are connection coefficients and make up the torsion tensor. In the final part \cref{E} we construct a curvature scalar that retains the antisymmetric part in a consistent way; the resulting invariant action encodes a gravitational theory in the background of $Q$ and $R$. We end in section \ref{C} with comments, discussion and outlook.

\section{Graded Poisson structure}
\label{A}

Graded geometry and super geometry are  popular in physics, primarily because they are convenient for dealing with manifest supersymmetry. The significance of the supersymmetry algebra was in particular noticed in \cite{Haag:1974qh} because it gives a chance to circumvent the Coleman-Mandula theorem \cite{Coleman:1967ad}, a powerful no-go theorem that severely constrains a relativistic quantum field theory to have only internal symmetries besides Poincar\'{e} symmetry. When a nilpotent differential operator is also part of the geometrical data, there are immediate applications in the context of BRST and BV-BFV quantization of theories,
 whose action is degenerate because of the gauge redundancies, and  in the context of AKSZ sigma models. There has also been a recent proposal for a formulation of T-duality in this framework, see \cite{Deser:2019pue}. This section aims at presenting the canonical Poisson structure of a specific graded manifold, together with a naturally associated differential operator. We will introduce the required underlying theory while trying to keep it accessible to physicists with expertise in theoretical gravitational physics. By the end of this  section, we will propose a very general deformed graded Poisson algebra, that can ultimately accommodate  the geometrical data needed for a gravitational action with stringy background fields.

\subsection{Canonical ($T^{*}[2]T[1]M, \omega, \text{Q}$)}
\label{can-tr}

Let us consider the graded $4d$-dimensional manifold given by the shifted cotangent of the shifted tangent space to a $d$-dimensional manifold $M$, $T^{*}[2]T[1]M \equiv \mathcal{M}$. The integer numbers in the square brackets refer to the degree of the coordinates of the corresponding space. According to Batchelor's theorem, the structure sheaf\footnote{Denoted as $\mathcal{O}_\mathcal{M}$, it can be thought as the algebra of functions on the graded manifold.} is isomorphic to the sheaf of sections of the exterior algebra of $E$, which is a vector bundle over $M$.  The graded manifold $\mathcal{M}$ admits a natural symplectic structure and thus a Darboux chart. The coordinates of the Darboux chart in a small neighborhood of a point $q\in M$, corresponding to a chart $\{ x^{i} \} $ on $M$, are degree $1$ $(\chi_{i})$ in the fiber of $T[1]M$, degree $1$ $( \theta^{i} ) $ for the fiber of $T^{*}[1]M$ and degree $2$ $( p_{i} ) $ in $T^{*}[2]M$. The degree goes along with the intrinsic parity of the structure sheaf in the following sense: coordinates of even degree commute with themselves, while those of odd degree anticommute. Denoting with $\upsilon(x), U(x)$ and $V(x)$ linear functions of $p, \chi, \theta$ respectively, and $f(x) \in C^{\infty}(M)$, the non-trivial Poisson brackets are given by:
\begin{equation}
\begin{matrix*}[l]
\{ p_{j}, x^{i} \} = \delta_{j}{}^{i}, & \, & \{\upsilon(x), f(x)\} = \upsilon^{i}(x) \partial_{i} f(x) ,\\
\{\chi_{i}, \theta^{j} \} = \delta_{i}{}^{j} = \{\theta^{j}, \chi_{i} \} , & \, & \{U(x),V(x)\} = \eta(U(x),V(x)),
\end{matrix*}
\label{can_sym}
\end{equation}
where $\eta := \delta_{i}{}^{j} \mathrm{d}\theta^{i} \wedge \mathrm{d}\chi_{j}$. Canonical transformations for this graded phase space represented by degree-$2$ maps preserve the total degree. Canonical transformations in which the degree is higher or lower than $2$ modify the grading. The most general degree 2 generators are
\begin{equation}
h = \upsilon^{i}(x) p_{i} + \frac{1}{2} N^{\alpha \beta}(x) \xi_{\alpha} \xi_{\beta}, \quad N^{\alpha \beta}(x) \in \Gamma(\Lambda^{2}\left(T^{*}[1]M \oplus T[1]M \right)).\label{c-trf} \end{equation}
$h$ acts on functions of the graded Poisson space via Poisson brackets to produce infinitesimal canonical transformations. The first term yields infinitesimal diffeomorphisms of $M$ while the second term induces $O(d,d)$ rotations on the elements of the vector space $T^{*}[1]M \oplus T[1]M$. A general excursion on infinitesimal canonical transformations is postponed to section \ref{g-syms}.

\bigskip

\emph{Remark}: Focusing on the space of the degree-$1$ coordinates, a degree-preserving canonical transformation  introduces what is commonly known in Generalized Geometry as the generalized metric $\mathcal{H}$ on $T^{*}[1]M \oplus T[1]M $. As a symmetric tensor on the fibers, in every coordinate patch it depends on a Riemannian metric $g$ on the base and on a $2$-form $B \in \Lambda^{2}T^{*}M \equiv \Omega^{2}(M)$. Globally the generalized metric characterizes the bundle with the pair $(g,H)$ for $g$ as before and $H$ a closed $3$-form. In the realm of Generalized Geometry, $\mathcal{H}$ is defined from $\eta$ and an involution $\tau \in \text{End}(T^{*}[1]M \oplus T[1]M$), $\tau^{2}= \text{id}$. In the graded symplectic setting the generating function (of type 1) for the canonical transformation $F(\tilde{\theta}, \theta)$ must be antisymmetric, $F(\tilde{\theta}, \theta) = - F(\theta, \tilde{\theta})$ in order to define an involution. One can easily verify that
\[
F(\tilde{\theta}, \theta) = \theta g(x)  \tilde{\theta} + \frac{1}{2} \tilde{\theta} B(x) \tilde{\theta} - \frac{1}{2} \theta B(x) \theta, \quad g(x) \in \text{Sym}(T^{*}M \otimes T^{*}M), B(x) \in \Omega^{2}(M)
\]
is the most general degree $2$ function with the required properties. In fact given the expression for the conjugate new coordinates, denoted with a tilde on top, and the old coordinates as functions of the known coordinates:
\[
\chi = \dfrac{\partial F}{\partial \theta}, \quad \tilde{\chi} = - \dfrac{\partial F}{\partial \tilde{\theta}},
\]
we can derive the following formula:
\[
\begin{pmatrix} \tilde{\chi} ,  & \tilde{\theta} \end{pmatrix} = \begin{pmatrix} \chi, & \theta \end{pmatrix} \begin{pmatrix} g^{-1}B & & g^{-1} \\ g-Bg^{-1}B  & & -Bg^{-1} \end{pmatrix} = \begin{pmatrix} \chi, & \theta \end{pmatrix} \mathcal{H} ,
\]
and the canonical Poisson brackets $\{ \tilde{\theta}^{j}, \tilde{\chi}_{i} \} = \delta^{j}{}_{i}$ while $\{ \tilde{\theta}, \tilde{\theta}\} = 0 = \{\tilde{\chi} , \tilde{\chi} \} $.

\bigskip

Our graded Poisson structure can be equipped with a further object: a homological vector field $\text{Q}$ compatible with the above Poisson structure, $\mathcal{L}_{\text{Q}} \omega =0$. Then $(T^{*}[2]T[1]M, \omega, \text{Q})$ becomes what is technically called a QP-manifold. Contrary to the ungraded case, in graded symplectic geometry every such vector field is Hamiltonian, under the mild condition that the degree of $\text{Q}$ and $\omega$ do not sum to zero. If $ \text{Q}$ has degree $1$, then by degree counting it is determined by a degree $3$ Hamiltonian $\Theta$ as 
\begin{equation} \text{Q}=\{ \cdot, \Theta \}. \label{QT} \end{equation}
The most general instance of $\Theta$ is
\begin{equation}
\Theta = \xi_{\alpha} \tilde{\rho}^{\alpha i}(x) p_{i} + \frac{1}{3!} C^{\alpha \beta \gamma}(x) \xi_{\alpha} \xi_{\beta} \xi_{\gamma} \, .
\label{hamil}
\end{equation}
In the expression for the Hamiltonian we use the shorthand notation $\xi_{\alpha} := \left( \chi_{i}, \theta^{i}\right)$, where $\alpha$, and all Greek indices from now on, are understood to take values from $1$ to $2d$; $C^{\alpha \beta \gamma}(x)$ is a section of the tensor space $\Lambda^{3} \left(T[1]M \oplus T^{*}[1]M \right)$. Then $\tilde{\rho}$ is a smooth map from the dual of $T[1]M \oplus T^{*}[1]M$, taken with respect to the bilinear and symmetric form there, to $T^{*}[2]M$. 
\[
\tilde{\rho}^{\alpha i} := \left( \rho \circ \eta \right)^{\alpha i}, \quad \text{for} \; \rho: T^{*}[1]M \oplus T[1]M \mapsto T^{*}[2]M.
\]

The homological vector field associated with \eqref{hamil} can be computed explicitly with the Poisson brackets \eqref{can_sym}: 
\[
\text{Q} = \xi_{\alpha} \partial_{k} \tilde{\rho}^{\alpha i}(x) p_{i} \dfrac{\partial}{\partial p_{k}} -\xi_{\alpha} \tilde{ \rho}^{\alpha i}(x) \dfrac{\partial}{\partial x^{i}}  + \tilde{\rho}^{\alpha i}(x) p_{i} \dfrac{\partial}{\partial \tilde{\xi}^{\alpha}} + \frac{1}{2} C^{\alpha \beta \gamma}(x) \xi_{\beta} \xi_{\gamma} \dfrac{\partial}{\partial \tilde{\xi}^{\alpha}} +\frac{1}{6} \partial_{k} C^{\alpha \beta \gamma}(x) \xi_{\alpha} \xi_{\beta} \xi_{\gamma} \dfrac{\partial}{\partial p_{k}} ,
\]
where $\tilde{\xi}^{\alpha}= \left(\theta^{i}, \chi_{i} \right)$ is the dual coordinate to $\xi_{\alpha}= \left( \chi_{i}, \theta^{i} \right)$. The dual is taken with the help of the metric $\eta$.

The (co)homological condition $\text{Q}^{2}=0$ can be rearranged using the definition \eqref{QT} and the graded Jacobi identity into:
\begin{equation}
\{ \Theta, \Theta \} = 0,
\label{MASTER}
\end{equation}
known as the classical master equation, which is highly non-trivial because of the even degree of the expression. This differential equation imposes restrictions on the possible functions $\tilde{\rho}(x)$ and $C(x)$ in the Hamiltonian for a given graded symplectic manifold. It is known \cite{Heller:2016abk} that \eqref{MASTER} gives rise to a system of three algebraic and differential equations: one constrains the image of $\tilde{\rho}$, a subset of $T^{*}[1]M \oplus T[1]M$, to be the kernel of $\rho$. Another one involves the tensors collectively incorporated in $C(x)$, which can be unpacked in $H_{ijk} \theta^{i} \theta^{j} \theta^{k}$, $f^{i}{}_{jk} \chi_{i} \theta^{j} \theta^{k} $, $Q^{ij}{}_{k} \chi_{i} \chi_{j} \theta^{k} $ and $R^{ijk} \chi_{i} \chi_{j} \chi_{k}$, in analogy to the stringy fluxes. The equation is hence untied in the Bianchi identities for $H$, $f$, $Q$ and $R$, which were firstly derived in the context of Generalized Geometry and algebroids in reference \cite{Blumenhagen:2012pc}, and were originally obtained by applying a T-duality on the Bianchi identity for $H$ in the seminal paper \cite{Shelton:2005cf}. The fluxes are locally given in terms of a $2$-form $B$ and a non-Poisson bivector $\Pi$, their most general coordinate expressions in a flat background being:
\begin{equation}
\begin{matrix*}[l]
H_{ijk} = \partial_{[i} B_{jk]}  \\
f^{i}{}_{jk} = [e_{j}, e_{k}]^{i} + \Pi^{im} H_{mjk}  \\
Q^{ij}{}_{k} = \partial_{k} \Pi^{ij} + \Pi^{il} \Pi^{jm} H_{lmk}  \\
R^{ijk} = 3 \Pi^{[ i \vert m} \partial_{m} \Pi^{\vert jk]} + \Pi^{il} \Pi^{jm} \Pi^{kn} H_{lmn}.
\end{matrix*}
\label{flu}
\end{equation}
In the second row, $f$ is based on the Lie bracket of gauge fields $e= e_{j}{}^{m}(x) \partial_{m}$. For our purposes we will resort to a less general definition in due course, namely $f$, $Q$ and $R$ will not depend on the $H$ flux.

The remaining equation retrieved from \eqref{MASTER} shows that, depending on $\rho$, some of the fluxes are identically zero. The canonical choice for $\Theta$ is met with the functions 
\begin{equation} \rho(\chi_{j}) = p_{j} , \quad \rho(\theta^{j}) = 0, \quad \quad C^{\alpha \beta \gamma} \xi_{\alpha} \xi_{\beta} \xi_{\gamma} = H_{ijk} \theta^{i} \theta^{j} \theta^{k}.
\label{std-rhoC}
\end{equation}

\subsection{Non-canonical $T^{*}[2]T[1]M$ via deformation}
Now we want to consider a different closed non-degenerate $2$-form $\omega^{\prime}$ on the same graded manifold. In the case of a regular non-graded symplectic manifold, Moser lemma \cite{Moser1965} guides us in the choice. For a given $\omega$, it allows us to take a whole $1$-parameter family of symplectic forms $\omega_{t}$ such that $\omega_{t} = \omega + F_{t}$, where $F_{0} = 0$. Moser lemma then states that there exists a $1$-parameter family of diffeomorphisms $\varphi_{t}$ relating $\omega_{t}$ to $\omega$ in the usual way:
\begin{equation}
\varphi_{t}^{*} \omega_{t} = \omega. 
\label{pull}
\end{equation}
The proof goes as follow. Although the isotopy $\varphi: \mathcal{M} \times [0,1] \mapsto \mathcal{M}$ could cease to be a flow because the composition law ($\varphi_{t} \circ \varphi_{s} = \varphi_{t+s}$) might not hold, it can still be regarded as a flow generated by a vector field $X_{t}$. Differentiating \eqref{pull} with respect to $t$, the candidate generating vector $X_{t}$ must satisfy the differential equation:
\[
\iota_{X_{t}} \omega_{t} = - \dfrac{\mathrm{d}}{\mathrm{d}t} A_{t} ,
\]
where $A_t$ is a local gauge field satisfying $F_{t} = \mathrm{d} A_{t}$. The above expression can be solved for $X_{t}$, since the symplectic form is non-degenerate. Moser lemma generalizes to the case of graded symplectic manifolds. Diffeomorphisms will be vector bundles isomorphisms covering morphism of the base manifold.

Note that the result of the lemma holds globally. Despite this, in this article the deformation will be characterized locally, as we are considering a patch for the neighborhood $U_{\text{q}}$ of a point q $\in M$, in which the symplectic form $\omega$ is written in Darboux coordinates. In due course we will point out that our statements are global. 

Given the above premises, let us deform $\omega$ by means of a linear, differentiable and invertible local map \begin{equation} \mathcal{E}(x): T^{*}[1]U_{\text{q}} \oplus T[1]U_{\text{q}} \rightarrow T^{*}[1]U_{\text{q}} \oplus T[1]U_{\text{q}}. \label{EE} \end{equation}
Then in the same neighborhood $U_{\text{q}}$ the new symplectic form $\omega' \equiv \omega_{1}$ coincides with the canonical form when pulled back by $\mathcal{E}(x)$ \eqref{EE}: 
\begin{align}
\omega^{\prime} & = \mathrm{d}x^{i} \wedge \mathrm{d} p_{i} + \mathrm{d} \left(  \xi_{\gamma} \left(\mathcal{E}(x)^{-1}\right)^{\gamma}{}_{\alpha}  \right) \eta^{\alpha \beta}\wedge \mathrm{d} \left(  \xi_{\delta} \left(\mathcal{E}(x)^{-1}\right)^{\delta}{}_{\beta} \right) , \notag \\
\; & = \mathrm{d} x^{i} \wedge \mathrm{d} p_{i} + \mathrm{d} \xi_{\gamma} \mathcal{G}^{\gamma \delta}(x) \wedge \mathrm{d} \xi_{\delta} - \mathrm{d} x^{i} \left[ \xi_{\gamma} \mathit{\Gamma}_{i \alpha}{}^{\gamma} \right] \mathcal{G}^{\alpha \delta} \wedge \mathrm{d} \xi_{\delta} + \mathrm{d}x^{k} \left[ \xi_{\gamma} \mathit{\Gamma}_{k \mu}{}^{\gamma} \mathcal{G}^{\mu \nu} \xi_{\delta} \mathit{\Gamma}_{l \nu}{}^{\delta} \right] \wedge \mathrm{d} x^{l} . \label{def_sym}
\end{align}
We define $\mathcal{G}^{\gamma \delta} \equiv \left( \mathcal{G}^{-1} \right)^{\gamma \delta}$ to be the inverse of a pseudo-Riemannian metric:
\begin{equation}
\mathcal{G}^{-1} = \mathcal{E}^{-1} \eta \mathcal{E}^{-T}.
\label{vielbein}
\end{equation}
Furthermore $\mathcal{E}$ \eqref{EE} is a local section for the frame bundle, where the local frames over q are $O(d,d)$ symmetric. Equivalently, $\mathcal{E}$ is a family of $1$-forms with values in $\mathfrak{o}(d,d)$. We also define $\mathit{\Gamma}_{i \beta}{}^{\alpha}$ to be 
\begin{equation}
\mathit{\Gamma}_{i \beta}{}^{\alpha} :=   \left( \mathcal{E}^{-1}\right)_{\beta}{}^{\delta} \partial_{i} \mathcal{E}_{\delta}{}^{\alpha}.
\label{coef-con}
\end{equation}
In the notation of \eqref{can_sym}, the local deformed graded Poisson brackets are now:
\begin{equation}
\begin{matrix*}[l]
\{p_{i}, x^{j}\} = \delta_{i}{}^{j}, & \, & \{ \upsilon(x), f(x) \} = \upsilon(x)^{i} \partial_{i} f(x),\\
\{\xi_{\alpha}, \xi_{\beta}\} = \mathcal{G}_{\alpha \beta}, & \, & \{ U(x), V(x)\} = \mathcal{G}(U,V), \\
\{ p_{i}, \xi_{\alpha} \} = \mathit{\Gamma}_{i \alpha}{}^{\beta} \xi_{\beta}, & \, & \{ \upsilon(x), U(x) \} = \nabla_{\upsilon} U(x) , \\
\{ p_{i}, p_{j} \} = \text{R}^{\alpha}{}_{\beta i j} \xi_{\alpha} \mathcal{G}^{\beta \gamma} \xi_{\gamma}, & \, & \{ \upsilon(x), \kappa(x) \} = [\upsilon(x), \kappa(x)]_{\text{Lie}} + \text{R}(\upsilon(x), \kappa(x)),
\end{matrix*}
\label{PB-new}
\end{equation}
where $\text{R}^{\alpha}{}_{\beta i j}$ is the curvature tensor of $\nabla$:
\begin{equation}
\text{R}^{\alpha}{}_{\beta i j} = \partial_{[i} \mathit{\Gamma}_{j] \beta}{}^{\alpha} + \mathit{\Gamma}_{[i \vert \delta}{}^{\alpha} \mathit{\Gamma}_{\vert j] \beta}{}^{\delta}.
\end{equation}
Metric, connection and curvature are $x$-dependent. However with the connection coefficients in \eqref{coef-con} the curvature tensor is identically zero:
\[
\text{R}^{\alpha}{}_{\beta ji} = \partial_{[j \vert} \mathcal{E}^{\alpha}{}_{\delta} \partial_{\vert i]} \left(\mathcal{E}^{-1} \right)^{\delta}{}_{\beta} + \mathcal{E}^{\alpha}{}_{\delta} \partial_{[ j \vert} \left(\mathcal{E}^{-1}\right)^{\delta}{}_{\gamma} \, \mathcal{E}^{\gamma}{}_{\mu} \partial_{ \vert i] } \left( \mathcal{E}^{-1}\right)^{\mu}{}_{\beta}= 0.
\] 
This aspect is due to the particular choice of the diffeomorphism, which is a redefinition of the degree-$1$ coordinates leaving the base invariant. One could have sought more general types of transformations, yielding additional $\mathrm{d}x \wedge \mathrm{d}x$ $2$-forms which by definition have the right properties of a curvature tensor (see \eqref{R-curv}). The current symplectic form in the chart discussed here is enough for our purposes. For the rest of the subsection $\text{R}^{\alpha}{}_{\beta ij}$ will nevertheless be kept explicit to stress the features of the construction.

Some considerations are in order: the connection is metric w.r.t.\,$\mathcal{G}$. This feature is implied by Jacobi identity of the Poisson bracket, for a pair of linear functions in $\chi, \theta$ and one linear function of $p$.
\begin{align}
\{\upsilon(x), \{ U(x), V(x) \}\} &= \{\{\upsilon(x), U(x)\}, V(x)\} + \{U(x) , \{ \upsilon(x), V(x)\}\},  \notag \\
\Rightarrow v(x) . \mathcal{G}(U,W)  & = \mathcal{G}(\nabla_{\upsilon(x)} U, W) + \mathcal{G}(U, \nabla_{\upsilon(x)} W). \label{metricityy}
\end{align}
The dot refers to the action of the derivation $\upsilon(x)$ on the $C^{\infty}(M)$-element that follows. \eqref{metricityy} is the condition for the metric to be covariantly constant, i.e.\,$\nabla_{\upsilon(x)} \mathcal{G} = 0$.

The differential Bianchi identity (or second Bianchi identity) for the curvature is itself an outcome of Jacobi identity. Let us call $\upsilon(x)$, $\kappa(x)$ and $\varepsilon(x)$ a triplet of linear functions in the momenta:
\begin{align}
0 & = \left\{ \upsilon(x),  \left\{ \kappa(x), \varepsilon(x) \right\} \right\} - \left\{ \left\{ \upsilon(x), \kappa(x) \right\}, \varepsilon(x) \right\} - \left\{ \kappa(x) , \left\{ \upsilon(x), \varepsilon(x) \right\}\right\}, \notag \\
\Rightarrow &  [\upsilon , [\kappa, \varepsilon]] - [[\upsilon, \kappa], \varepsilon] - [\kappa, [\upsilon, \varepsilon]] = 0,  \; \; \; \; \; \; \; \text{Jacobi identity for the Lie bracket} \notag\\
\, & \nabla_{\upsilon} \text{R}^{\alpha}{}_{\beta}(\kappa, \varepsilon) + \nabla_{\varepsilon} \text{R}^{\alpha}{}_{\beta}(\upsilon, \kappa) + \nabla_{\kappa} \text{R}^{\alpha}{}_{\beta}(\upsilon, \varepsilon) + \text{R}^{\alpha}{}_{\beta}(\upsilon, [\kappa, \varepsilon]) - \text{R}^{\alpha}{}_{\beta}([\upsilon, \kappa], \varepsilon) - \text{R}^{\alpha}{}_{\beta}(\kappa, [\upsilon, \varepsilon])= 0.
\end{align}
In the last row the second Bianchi identity is reproduced in the form
\[
\left(\nabla_{\upsilon} \text{R} \right)(\kappa, \epsilon) + \text{R}(\text{T}(\upsilon, \kappa), \epsilon) + \text{cyclic} = 0. \quad \text{T}(\upsilon, \kappa) := \nabla_{\upsilon} \kappa - \nabla_{\kappa} \upsilon - [\upsilon, \kappa].
\]

Degree counting constrains the possible combinations of functions in the Poisson brackets. The only possibility we are left to check is $\{ V, \{ \upsilon, \kappa\}\}$. The Jacobi identity is verified if $\text{R}$ is bound to be the curvature tensor.
\begin{align}
\{V, \{ \upsilon, \kappa\}\} - \{\{V, \upsilon\}, \kappa\} - \{ \upsilon, \{ V, \kappa \}\} = 0, \notag \\
\Rightarrow \nabla_{\upsilon} \nabla_{\kappa} U - \nabla_{\kappa} \nabla_{\upsilon} U - \nabla_{[\upsilon, \kappa]} U - \text{R}(\upsilon, \kappa) U = 0. \label{R-curv}
\end{align}

A compatible differentiable structure (Q-structure) on $(T^{*}[2]T[1]M, \omega^{\prime})$ is still formally given by the same Hamiltonian function and vector field of the canonical case. To verify the classical master equation with the new Poisson structure \eqref{PB-new}, the maps $\tilde{\rho}$ and $C$ shall subsequently respect different conditions than in the previous case. Let us uncover a bit more what this means in the completely general setup, when $\text{R}^{\alpha}{}_{\beta ij}$ is not set to its null value:
\begin{equation}
\begin{matrix*}[l]
\tilde{\rho}^{\gamma j } \mathcal{G}_{\gamma \alpha} \tilde{\rho}^{\alpha i} = 0 ,
\\
\tilde{\rho}^{\gamma i} \mathcal{G}_{\gamma \sigma} C^{\sigma \alpha \beta} +2 \tilde{\rho}^{\beta j } \left( \mathit{\Gamma}_{j \gamma}{}^{ \alpha } \tilde{\rho}^{\gamma i} + \partial_{j} \tilde{\rho}^{ \alpha i} \right) =0 , 
\\
 \dfrac{1}{3} \tilde{\rho}^{\alpha i} \left( \partial_{i} C^{\beta \gamma \delta} + 3\mathit{\Gamma}_{i \sigma}{}^{\beta} C^{\sigma \gamma \delta} \right) + \dfrac{1}{4} C^{\alpha \beta \sigma} \mathcal{G}_{\sigma \mu} C^{\mu \gamma \delta} & =-\text{R}^{\gamma \delta}{}_{ij} \tilde{\rho}^{\alpha i} \tilde{\rho}^{\beta j}. 
\end{matrix*}
\label{first-master}\end{equation}
All the non-contracted Greek indices are understood to be antisymmetrized. The first algebraic equation, if $\mathcal{G}$ is locally diagonalized by some vielbeins $\mathcal{F}$, is a rewriting of the claim that the image of $\mathcal{F}^{*} \circ \tilde{\rho}^{*}$, a subset of $T^{*}[1]M \oplus T[1]M$, is the kernel of $\tilde{\rho} \circ \mathcal{F}$, i.e.\,$\eta(\mathcal{F}^{*} \circ \tilde{\rho}^{*}, \mathcal{F}^{*} \circ\tilde{\rho}^{*}) = 0$; as we will see in the next section, this appoints the underlying Courant algebroid to be exact (see around \eqref{seq}). The other equations are covariant versions of what discussed below \eqref{MASTER}. The second row tells us how the given metric and its connection $\mathit{\Gamma}$ must be related to the tensors $C$ and the linear map $\tilde{\rho}$. Then the third line displays the generalization of the Bianchi identities for the fluxes $C$: usually in differential form they will be written with a twisted de Rham differential, but here they are expressed with the covariant derivative and the inner product with $C$ taken with the $G$ metric. The non-zero curvature $\text{R}$ of the connection $\nabla$ obstructs the cochain complex. 

In the present situation, the curvature is locally zero, and the setup is comparable to having Riemann normal coordinates in standard differential geometry. Thus the curved metric can be ``off-diagonalized'' into $\eta$, by means of the frame fields $\mathcal{F}$, and it is possible to seek as metric connection an antisymmetric connection (i.e.\,with non-zero torsion). If $\mathcal{G}$ and $\mathcal{F}$ are fixed in the first algebraic equation, its meaning is thus clear, and it can be explicitly solved for $\tilde{\rho}$ with the aforementioned properties. Then an available solution to the third equation can always be found in $C \equiv 0$\footnote{Focusing on a Hamiltonian \eqref{hamil} that consists only of the first summand  leads to the same conclusion: the third equation in \eqref{first-master} is $0 \equiv 0$ and in the second equation the first term drops out.}.  
Then the second differential equation in \eqref{first-master} reduces to $\nabla^{\alpha} \tilde{\rho}^{\beta i} = 0$ and determines $\mathit{\Gamma}$ in terms of derivatives of the vielbein and of $\tilde{\rho}$, and viceversa. For \eqref{coef-con} this equation is immediately true. 

Notice that we can anyway consider generalized closed $3$-forms built from the local antisymmetric $2$-tensor gauge fields. These gauge fields are respectively the Kalb-Ramond field $B$, arbitrary and non-constant, and for the other tensors in the T- duality chain the bivector $\Pi$. They can actually be incorporated through the deformation (in the vielbein and ultimately the connection). Instead if we were deforming the Poisson bracket in other ways which yielded non-zero curvature in the Poisson bracket of two functions of $p$, the components of $C$ could not be set all simultaneously to zero. Focusing on $B$ and $\Pi$ and their field strengths alone would not be possible, as cohomologically non-trivial $H$, $f$, $Q$ or $R$ can be present as well.

In the case where the local differentiable and invertible map that pulls back $\omega'$ into $\omega$, expressed in the Darboux chart, is known (it can be a vielbein $\mathcal{E}(x)$ like here, but also something else), an effortless expedient to obtain the solutions $\tilde{\rho}$ and $C$ to the classical master equation is at hand. It consists in exploiting the local map to pull back or push forward an already known solution for the canonical case. Hence in our case the following transformations applied to $\tilde{\rho}$ and $C$ give consistent tensors that can be used in the Hamiltonian $\Theta$: 
\begin{equation}
\tilde{\rho}^{\alpha i} \mapsto \left(\mathcal{E}^{-1}\right)^{\alpha}{}_{\beta} \tilde{\rho}^{\beta i}=: \underline{\tilde{\rho}}^{\alpha i}, \quad  C^{\alpha \beta \gamma} \mapsto \left(\Lambda^{3} \mathcal{E}^{*} C \right)^{\alpha \beta \gamma} =: \underline{C}^{\alpha \beta \gamma} .
\label{anchor}
\end{equation}
We will hence use the canonical maps in \eqref{std-rhoC} and transform them by means of the vielbein map $\mathcal{E}$. We will omit the subscript from now on, as this shall not cause confusion. Moreover we will take $C=0$ as already announced. 

\subsection{Gauge symmetries}
\label{g-syms}

In this part we pause from the main exposition to spend a few words on the symmetries of our model, expanding the discussion around formula \eqref{c-trf}. A graded Poisson algebra of any degree and a corresponding Hamiltonian can be invariant under some transformations with infinitesimal generators $\delta_{\alpha}$, acting on the coordinates via Poisson brackets. Since $\delta_{\alpha}$ is a graded derivation for the Poisson algebra, the infinitesimal transformations it induces are canonical: the Poisson brackets remain invariant especially because their associativity property is preserved. A first distinction between these transformations can be made following the degree: we can distinguish between degree-preserving transformations and degree changing ones. In the case under investigation here of an odd Hamiltonian $\Theta$ of degree $3$, and even Poisson brackets $\vert \{,\}\vert = -2$, the transformed Hamiltonian $\Theta + \delta_{\alpha} \Theta$ will also solve the classical master equation to first order in $\delta_{\alpha}$ (higher orders are neglected because the transformation is infinitesimal).

Infinitesimal canonical transformations of a Poisson algebra of degree $2$ were classified by Roytenberg in \cite{Roytenberg:2002nu} for the degree-preserving case. He found that they generate diffeomorphisms and $O(d,d)$-transformations, \eqref{c-trf}. In particular $B \in \Lambda^{2}T[1]M$ and $\Pi \in \Lambda^{2}T^{*}[1]M$ belong to the Lie algebra $\mathfrak{o}(d,d)$.

When a $\text{Q}$-structure is present, one may wonder whether the algebraic objects obtained via derived brackets can share the same symmetry with the original Poisson algebra. This boils down to ask if the Hamiltonian can be invariant under the symmetry, $\delta_{\alpha} \Theta = 0$, or equivalently if $\alpha$ can be closed with respect to the differential $\text{Q}$,  $\text{Q}\alpha =0$. This implies that in a locally contractible open set the function $\alpha$ must be $\text{Q}$-exact and hence given by 
\[
\alpha = \{ \varrho, \Theta\} , \quad \quad \vert \varrho \vert + \vert \Theta\vert + \vert \{\cdot, \cdot\}\vert = \vert \alpha \vert,
\] 
since then $\delta_{\alpha} \Theta = \{\alpha, \Theta\} = \{\{\varrho, \Theta\}, \Theta\} = \dfrac{1}{2}\{ \varrho, \{\Theta, \Theta\}\}=0$, irregardless of the even or odd degree of the functions involved. For the sake of our purposes, we are interested in a degree-$2$ function $\alpha$ of the structure sheaf of the degree $2$ dg-symplectic manifold $T^{*}[2]T[1]M$. Therefore we want to let $\Theta$ act on a degree $1$ function, $\varrho = \varrho^{\mu}(x) \xi_{\mu}$. If the Hamiltonian is given by
\[
\Theta = \xi_{\alpha} \tilde{\rho}^{\alpha i} p_{i},
\]
and using the deformed Poisson brackets \eqref{PB-new} we obtain:
\begin{align*}
\{ \varrho, \Theta\} = \varrho^{\mu}(x) \mathcal{G}_{\mu \alpha} \tilde{\rho}^{\alpha i}p_{i} + \xi_{\alpha} \tilde{\rho}^{\alpha i} \nabla_{i} \left(\varrho^{\mu}(x) \xi_{\mu} \right).
\end{align*}
The first term generates ``Q-exact'' diffeomorphisms, while the second is responsible of ``Q-exact'' $O(d,d)$ transformations. Anyway, we would like to restrict our considerations to the latter ones. In this way we are representing the gauge symmetries of the endomorphisms of $T^{*}[1]M \oplus T[1]M$, of $B \in \Lambda^{2}T[1]M$ and $\Pi \in \Lambda^{2}T^{*}[1]M$ in the graded geometrical setting. We impose that only degree $2$ functions quadratic in the $\xi$ can appear: this occurs when $\varrho$ is mapped to the null vector field by the anchor:
\begin{equation}
\rho(\varrho) = 0. \label{var0} 
\end{equation}

In conclusion, for future reference we are going to need this particular $\alpha$: 
\begin{equation}
\alpha^{\beta \gamma} = \tilde{\rho}^{\beta i} \nabla_{i} \varrho^{\gamma}(x),
\label{qex}
\end{equation}
with $\varrho$ subjected to the above \eqref{var0}. In particular, in the next sections we are going to check invariance under the gauge symmetries of the $\Pi$ field.

\subsection{Deformation and open-closed string relation} 
\label{ex}
For our purposes we will assign to the {}local invertible differentiable map for the degree $1$ coordinates $\mathcal{E}(x)$, the block matrix expression\footnote{The inverse of the local vielbein $\mathcal{E}(x)$ \eqref{trafo} will also be essential:
\begin{equation}
\mathcal{E}^{-1} = \begin{pmatrix} \frac{g}{2}^{-1}\left(g-B\right) & -\frac{g}{2}^{-1} \\ +\frac{G}{2} & \left(g-B\right)\frac{g}{2}^{-1}  \end{pmatrix}.
\label{trafo_inverse}
\end{equation}}:
\begin{equation}
\mathcal{E} = \begin{pmatrix} \mathbb{1} &  \left(g-B\right)^{-1} \\ - \left(g+B\right) & \mathbb{1} \end{pmatrix}.
\label{trafo}
\end{equation}
The metric $g$ is Riemannian, while the $2$-form $B$, seen as a smooth linear map $B : \Gamma(TM) \mapsto \Gamma(T^{*}M)$, is invertible. In the upper right $d$-dimensional block we invoke the {}open-closed string relation
\begin{equation}
\left(g-B \right)^{-1} =: G^{-1} - \Pi ,
\label{Sei-Wit}
\end{equation}
where $G^{-1} \in \vee^{2} T^{*}[1]M$ and $\Pi \in \Lambda^{2} T^{*}[1]M$ are non-degenerate tensors that can be separately written in terms of $g$ and $ B$ as:
\begin{equation}
G^{-1} := \left(g+B\right)^{-1} g \left(g-B\right)^{-1}, \quad \Pi := -\left(g+B \right)^{-1} B \left(g-B \right)^{-1}.
\label{Open_metric}
\end{equation}
In particular $G^{-1}$ is the inverse of a Riemannian metric. Equation \eqref{Sei-Wit} is the closed-open string relation \cite{Seiberg:1999vs} (in units where $2\pi \alpha' \equiv 1$): in that reference it was obtained comparing the open string propagator at the boundary of the cylinder, with the closed string propagator extrapolated for points in the interior of the cylinder. $G$ is the metric as seen by the open string, opposed to the metric $g$ seen by the closed string. In this context we do not need a deeper understanding of closed and open string theory but the relation between the quadruplet of tensors ($g, B, G, \Pi$) stated by \eqref{Sei-Wit}. In our related work \cite{Boffo:2020vqx} we analyzed a similar generalized vielbein, with the upper right block put to zero. The vielbein used here returns a class of block diagonal metrics, i.e. 
\begin{equation}
\mathcal{G} = \begin{pmatrix} -2g & 0 \\ 0 & 2G^{-1} \end{pmatrix}. 
\label{DIAG}
\end{equation}
The group $O(d) \times O(d)$ admits an action on $\mathcal{G}$; in the canonical setting $\eta$ is invariant under the larger group $O(d,d)$. Notice that, due to the opposite sign, $\mathcal{G}$ is Lorentzian of signature $(-d, d)$, as the $O(d,d)$-invariant pairing is.

Subsequently the symbols for the connection are as well functions of the fields $g,B, G^{-1}$ and $\Pi$, and they satisfy \eqref{metricityy}. $\mathcal{E}^{-1}$ is given in the footnote, \eqref{trafo_inverse}.
\begin{equation}
\begin{matrix*}[l]
\mathit{\Gamma}_{ i j}{}^{k} = \frac{1}{2} g^{km} \partial_{i} \left( g + B \right)_{mj} , & \; &
\mathit{\Gamma}_{i}{}^{jk} = \frac{1}{2} g^{k l} \left( g-B \right)_{lm} \partial_{i} \left( g- B \right)^{mj} ,\\
\; & \; & \; \\
\mathit{\Gamma}_{ijk} = -\frac{1}{2} \left( g-B \right)_{kl} g^{l m} \partial_{i} \left( g + B \right)_{mj},
 & \; &
\mathit{\Gamma}_{i \; \; k}^{\; j} = \frac{1}{2} G_{km} \partial_{i} \left( g- B \right)^{mj} ,
\end{matrix*}
\label{matrix-conn}
\end{equation}
Upper indices denote the (component of) the inverse of the full expression, i.e. $g^{ij} := \left(g^{-1} \right)^{ij}$ and $\left(g-B\right)^{ij} :=  \left(\left(g- B \right)^{-1}\right)^{ij}$.

The new coordinates obtained composing with \eqref{trafo} on the right deserve further attention. 
Introducing $B(x) \in \Lambda^{2} T[1]M$ and modifying the basis as $ \chi_{j} \mapsto \chi_{j} - \theta^{i}\left(g_{ij}(x) +B_{ij}(x)\right)$ introduces the gauge potential for the stringy $H$-flux, and a metric $g(x)$. Remarkably, a regular metric for tangent vectors is already directly present in the generalized vielbein itself, rather than being implemented via $d$-dimensional vielbeins, as sometimes happens. The relative sign between $g$ and $B$ is only a matter of convenience; most importantly, these two tensors are demanded to appear on the same footing (without different prefactors). Eventually this configuration is summed with its inverse (with reverted sign): $\theta^{j} \mapsto \chi_{i} \left( g- B \right)^{ij} + \theta^{j} $. The change of sign assures that $\mathcal{E}$ is invertible and hence an isomorphism of the Poisson algebra. The bivector $\Pi \in \Lambda^{2}T^{*}[1]M$ will source $Q \in \Lambda^{2}TM \otimes T^{*}M$ and $R \in \mathfrak{X}^{3}(M)$, in the same way as $B$ is the gauge field for a $3$-form $H$. Notice that the deformation of $\omega$, carried by the vielbein \eqref{trafo} based on $g$ and $B$, hints at the abelian $1$-gerbe structure of the related Courant algebroid (treated in \cref{CAA}). That is also the reason why we do not look at the most general Hamiltonian function (i.e.\,the $C^{\alpha \beta \gamma}(x)$ tensors are set to zero): the $H$-flux of the T-duality chain will already appear directly in the derived Courant algebroid bracket thanks to the new graded form $\omega'$.

Let us replicate for future reference the deformed Poisson brackets we are going to analyze in the course of the article:
\begin{equation}
\begin{matrix*}[l]
\{p_{j}, x^{i}\}'=\delta_{j}^{i}, & \{ \upsilon(x) , f(x) \}' = \upsilon(x). f(x),\\
\{\xi_{\alpha}, \xi_{\beta}\}' = \begin{pmatrix} -2g & 0 \\ 0 & 2 G^{-1} \end{pmatrix}, & \{ U(x), V(x) \}'= -2g(X,Y) + 2G^{-1}(\gamma, \sigma), \\
\{p_{i}, \xi_{\alpha}\}'= \mathit{\Gamma}_{i \alpha}{}^{\beta} \xi_{\beta} \; \eqref{matrix-conn}, & \{\upsilon(x), U(x) \}' = \nabla_{\upsilon} U(x), \\
\{p_{i}, p_{j}\}' = 0, & \{ \upsilon(x), \varkappa(x)\}' = [\upsilon, \varkappa]_{\text{Lie}}.
\end{matrix*}
\label{def-expl}
\end{equation}

For what concerns the explicit shape of the Hamiltonian $\Theta = \xi_{\alpha} \tilde{\rho}^{\alpha i} p_{i}$, substitution of \eqref{trafo} into \eqref{anchor}, where for the canonical graded Poisson algebra we adopt the standard projector $\rho_{\beta}{}^{i} = \left( \delta_{k}{}^{i}, 0 \right)$, returns 
\begin{equation}
\rho_{\alpha}{}^{i} = \left( \delta_{k}{}^{i}  ,\left(g+B\right)^{ki}\right), \quad \tilde{\rho}^{\alpha i} \equiv \mathcal{G}^{\alpha \beta} \rho_{\beta}{}^{i}= \left( -\dfrac{g}{2}^{ki}, \left[\left(g-B\right)\dfrac{g}{2}^{-1}\right]_{k}^{i} \right).
\label{anch-def}
\end{equation}
On the linear function $U = X + \sigma \in \Gamma(T^{*}[1]M \oplus T[1]M)$ it acts in this way:
\begin{equation}
\rho(U) = X + \left(g+B\right)^{-1}(\sigma) .
\label{dual-anchor}
\end{equation}
The Hamiltonian eventually becomes
\begin{equation}
\Theta = -\chi_{l} \frac{g}{2}^{li} p_{i} + \theta^{l} \left(g-B\right)_{lm} \frac{g}{2}^{mi} p_{i} \, .
\label{HAM}
\end{equation}
Checking whether the Poisson bracket of $\Theta$ \eqref{HAM} with itself closes to zero is easy once the Hamiltonian is constructed with \eqref{dual-anchor}. In conclusion, the graded Poisson algebra defined by \eqref{PB-new}, where the metric is \eqref{DIAG}, the connection coefficients are given in \eqref{matrix-conn} and the curvature is zero, together with $\Theta$ \eqref{HAM} yields a consistent QP-manifold.

\section{Courant algebroids and graded Poisson structures}
\label{CAA}

In this section we switch to Courant algebroids and generalized tangent bundles, since we would like to exploit tools of generalized differential geometry in our investigation. Again, we will try to present these mathematical concepts in a user-friendly way. Standard Courant algebroids were shown to be in direct relation with the $2$-graded QP-manifolds just illustrated. We are now going to first unveil the necessary information about Courant algebroids and then review the correspondence. 

\subsection{Correspondence between $2$-graded QP-manifolds and exact Courant algebroids}

It was shown in \cite{Roytenberg:2002nu} that a graded QP-manifold of degree $2$ with canonical symplectic structure corresponds to a Courant algebroid  on the bundle $T^{*}[1]M \oplus T[1]M \cong TM \oplus T^{*}M$. A Courant algebroid (in short CA) is a higher algebraic structure which consists of a vector or principal bundle endowed with a bracket, a surjective map from the total space to the tangent bundle and a symmetric bilinear form, so that some integrability properties are ensured. A CA on the tangent bundle is equivalent to a Lie algebroid, and over each point of the base space to a Lie algebra. Although the cases in which the construction of a Courant algebroid applies are much more general, for our purposes we will consider just \emph{exact} CAs on $E$. A CA on a vector bundle $E$ is exact when the short exact sequence
\begin{equation}
0 \rightarrow T^{*}M \xrightarrow{j} E \xrightarrow{\rho} TM \rightarrow 0 
\label{seq}
\end{equation}
holds. Every $E$ satisfying \eqref{seq} is hence isomorphic to $TM \oplus T^{*}M$ by means of a splitting of the sequence, and their algebraic structures intertwine.
\begin{defn} A \emph{Courant algebroid} is a quadruple $(E , \rho , [\cdot, \cdot ] , \langle \cdot, \cdot \rangle)$, where $\rho : E \rightarrow TM$ is the \emph{anchor} map, $[\cdot, \cdot] : \Gamma(E) \times \Gamma(E) \rightarrow \Gamma(E)$ is a bracket on sections and $\langle \cdot , \cdot \rangle : \Gamma(E) \vee \Gamma(E) \rightarrow C^{\infty}(M)$ is a fiber-wise symmetric bilinear form. The following minimal set of conditions must hold:
\begin{enumerate}
\item $[U, [V, W]] = [[U,V], W] + [ V, [U,W]] \quad \quad \quad \text{(left Leibniz identity for \,}[\cdot, \cdot]$),
\item $ \frac{1}{2} \rho(U) \langle V, V \rangle = \left\langle [U,V] , V \right\rangle \quad \quad \quad \text{(the pairing is invariant under the adjoint action),}$
\item $ \frac{1}{2} \rho(U) \langle V,V \rangle = \left\langle [V,V] , U \right\rangle \quad \quad \quad \text{(behaviour of the symmetric part of the bracket),}$
\end{enumerate}
for $U,V,W \in \Gamma(E)$. Notice furthermore that the anchor map is subjected to be actually $\rho \in \text{Hom}(E, TM)$ from the axioms. 
\label{defnCA}
\end{defn}
It is worth pointing out that the extended tangent bundle $TM \oplus T^{*}M$ has an abelian $1$-gerbe structure. The open sets are patched together through transition functions depending on a $2$-form $B$. In the non-empty intersection of open sets $U_{i} \cap U_{j}$,
\[
B_{(i)} = B_{(j)} - \mathrm{d} \lambda_{(ij)}.
\]
The exact $2$-form $\mathrm{d} \lambda$ must fulfill the cocycle condition: 
\[ \lambda_{(ij)} + \lambda_{(jk)} + \lambda_{(ki)} = c^{-1}_{(ijk)}\mathrm{d} c_{(ijk)}
\]
in the three overlapping patches $U_{i} \cap U_{j}  \cap U_{k} \neq \{0\}$, with $c_{(ijk)}$ holomorphic function with values in $U(1)$. $B$ is also invariant under $\lambda \mapsto \lambda + \mathrm{d}f$, $f \in C^{\infty}(M)$. Exact CAs are classified by the third cohomology class $H^{3}(M, \mathbb{R})$ \cite{Severa:2017oew}.

To be convinced that a QP-manifold $T^{*}[2]T[1]M$ encodes the same data as a CA on $TM \oplus T^{*}M$, the clue is to compute the master equation $\{ \Theta, \Theta \} = 0$ for the Hamiltonian given in \eqref{hamil}. It yields the set of defining conditions for a Courant algebroid of \cref{defnCA}. If $U(x), V(x) \in \Gamma(E)$ are given by $U^{\alpha}(x) \xi_{\alpha} = X^{i}(x) \chi_{i} + \sigma_{i}(x) \theta^{i},  V^{\alpha}(x) \xi_{\alpha} = Y^{j}(x) \chi_{j} + \kappa_{j}(x) \theta^{j}$, from the equivalence with the canonical QP-manifold ($T^{*}[2]T[1]M, \omega$  \eqref{can_sym}, $\Theta$ \eqref{hamil} with $\rho_{\alpha}{}^{i} = ( \delta_{k}{}^{i}, 0)$ and $C = H_{ijk} \theta^{i} \theta^{j} \theta^{k}$) one retains as natural bracket for the CA the so called Dorfman bracket,
\[
[U,V]_{\text{D}} := [X, Y] + \mathcal{L}_{X} \kappa - \iota_{Y} \mathrm{d} \sigma ,
\] 
twisted by $H$:
\[
\{\{U, \Theta\}, V \} = [U,V]_{H} := [U,V]_{\text{D}} + H(X,Y),  \quad H(X,Y) \in \Gamma(T^{*}M).
\]
This expression shows that the twisted Dorfman bracket is the derived bracket for the canonical graded Poisson algebra of degree-$1$ functions.

\subsection{Courant algebroid for a non-canonical QP-manifold}

So far we commented upon the relation between the Courant algebroid (CA) $(TM \oplus T^{*}M, \rho, [\cdot, \cdot], \langle \cdot, \cdot \rangle)$ and the associated canonical QP-manifold $(T^{*}[2]T[1]M, \omega, \Theta)$. Now we want to unveil the expressions of the CA objects for the non-canonical QP-manifold of our model instead. As already announced, in the Hamiltonian for the canonical case we resort to the identity map for $\rho$, while the tensors $C$ are set to zero. We will also present the results for the example of subsection \ref{ex} along with those for a generic vielbein $\mathcal{E}$. This should not create confusion but rather help to focus more immediately on the main outcomes.

The pairing $\langle \cdot, \cdot \rangle^{\prime}$ for the Courant algebroid is easily retrieved directly from the Poisson bracket between $U(x)$ and $V(x)$ linear functions in $\xi$:
\[
\{ U, V \} = \begin{pmatrix} -2g(X,Y) & 0 \\ 0 & 2G^{-1}(\sigma, \kappa)  \end{pmatrix} = : \langle U, V \rangle^{\prime}.
\]
The Dorfman bracket is generalized via derived bracket with the Hamiltonian in \eqref{HAM}, $\{ \{ \{ U, \Theta \} , V \},\\ W \} = \left\langle [U, V]^{\prime}_{\text{D}}, W \right\rangle^{\prime}$, with $W^{\alpha}(x) \xi_{\alpha} = Z^{k}(x) \chi_{k} + \zeta_{k}(x) \theta^{k}$ and $U,V$ as before, to the following:
\begin{align}
\langle[U,V]^{\prime}_{\text{D}}, W \rangle^{\prime} &=  \left\langle \nabla_{\rho(U)} V - \nabla_{\rho(V)} U , W \right\rangle^{\prime} + \left\langle \nabla_{\rho(W)} U, V \right\rangle^{\prime} \label{DORF}\\
\, & = \iota_{[\rho(U),\rho(V)]} \left( \zeta - \left(g-B\right)(Z)\right) + \iota_{[\rho(V),\rho(W)]} \left( \sigma - \left(g-B\right)(X)\right) \notag \\
\, & \; \; \;  - \iota_{[\rho(U),\rho(W)]} \left( \kappa - \left(g-B \right)(Y) \right) +  \rho(U) \left\langle \big(\kappa -\left(g-B \right)\left( Y \right) \big), \rho(W)\right\rangle  \notag \\
\, & \; \; \;  - \rho(V) \left\langle \big(\sigma -\left( g-B\right)\left( X \right)\big), \rho(W)\right\rangle + \rho(V)^{i} \rho (W). \big(\sigma -\left(g-B\right)\left(X\right) \big)_{i} , \label{comp_2}
\end{align}
where, in accordance with \eqref{dual-anchor}, the anchored vectors are 
\[
\rho(U) = X + \left(g+B \right)^{-1}(\sigma)
\] 
and \eqref{comp_2} displays the particular deformed graded Poisson structure under consideration from the general formula \eqref{DORF}. It might not be evident from the expression \eqref{comp_2}, but the bracket relies just on global objects: the anchor map, the Riemannian metric $g$ and $\mathrm{d}_{\rho}B \equiv H_{\rho} \in \Lambda^{3}(T^{*}M)$. The anchor-dependent differential $\mathrm{d}_{\rho}$ will be discussed in the next section \ref{d-defn}.

Notice also that part of expression \eqref{comp_2} for the bracket can be rearranged so to form again the fully contracted standard Dorfman bracket, i.e.\,$\langle [X,Y], \zeta \rangle + \langle \mathcal{L}_{X} \kappa - \iota_{Y} d \sigma, Z \rangle$, as one would expect when setting $g, B = 0$ in the deformation $\mathcal{E}$, since in that limit it reduces to the identity.

As a further consistency check, it is straightforward to directly prove that the new bracket $[\cdot, \cdot ]'_{\text{D}} $, the new block diagonal pairing and the new anchor $\rho$ respect the axioms in \cref{defnCA}. They are retrieved thanks to associativity of the deformed Poisson brackets, the master equation and Leibniz rule for the derivation $\text{Q}$. See also \cite{Heller:2016abk} for details.

Let us point out that in the Generalized Geometry description the generalized vielbeins $\mathcal{E}(x)$ are local sections of the $O(d,d)$ frame bundle. They can be extended on the overlapping patches: the transition functions are orthogonal transformations (which comprise diffeomorphisms, $B$ transforms and transformations with a bivector). These are canonical transformations of the underlying graded Poisson algebra, see the overview of section \ref{can-tr} and equation \eqref{c-trf}. Every two set of vielbeins will differ by an exact $2$-form: this is a feature of the underlying $1$-gerbe structure of the generalized tangent bundle. More importantly though, to get a CA on $TM \oplus T^{*}M$ one needs also a Hamiltonian function that can solve the classical master equation: different anchor maps and tensors $C$ can be good solutions of \eqref{first-master}. In this scenario, we will end up with a different representative of the third cohomology class of the Courant algebroid, or even with other tensors or multivectors involved. Nevertheless our results in the next sections will be affected just by addition of the possibly non-zero $C$-tensors to the derived connection \eqref{THM}.

\section{Differential geometry of $TM \oplus T^{*}M$}
\label{B}

In the upcoming section we review a bit of differential geometry of $TM \oplus T^{*}M \cong T^{*}[1]M \oplus T[1]M$. The material we cover is well established in the literature, but sometimes we complement it with some less standard notions. For example we will suggest a generalized version of the Lie bracket. Then, inspired by the expression of the deformed Dorfman bracket \eqref{DORF} and \eqref{comp_2}, and by subtraction of this generalization of a Lie bracket, we will be able to extract another metric connection (w.r.t.\,$\mathcal{G}$) on $E$-sections, with non-zero curvature in some sense yet to be defined. This fact is rather crucial. Indeed in the way we incorporated the gauge fields $B$ and $\Pi$ in the beginning we could just get the algebra of functions for a particle in a flat background, coupled to $B$ and $\Pi$ with the connection symbols \eqref{matrix-conn}. These fields were not dynamical themselves, since the curvature was identically zero. 

With the new connection, and by suggesting an invariant scalar of the curvature tensor fully projected into some subbundles, we will eventually obtain an action for a theory of gravity. Comments on its relation to the action of compactified string theory with $Q$ and $R$ fluxes, also known as the NS-NS action in a non-geometric frame \cite{Blumenhagen:2013aia}, will be postponed to the discussion \ref{C}. We already mentioned that the fluxes in the T-dual chain have found their description within the ordinary QP-manifold $(T^{*}[2]T[1]M, \omega, \Theta)$ as the $C^{\alpha \beta \gamma}(x)$ tensors in the Hamiltonian, that twist the CA bracket. In fact their nature is clear from the following piece of the derived bracket: 
\begin{align*}
\left\{ \left\{ \left\{ U, \frac{1}{3!}C^{\alpha \beta \gamma}(x) \xi_{\alpha} \xi_{\beta} \xi_{\gamma} \right\} , V\right\} , W \right\} =& \,  H(X,Y,Z) + f(X,Y,\zeta) + f(X,\kappa,Z) + f(\sigma,Y,Z) \\
\, & + Q(X,\kappa,\zeta) + Q(\sigma,\kappa,Z) + Q(\sigma, Y,\zeta) + R(\sigma, \kappa, \zeta).
\end{align*}
It was already pointed out in the previous section that they satisfy the Bianchi identities, in accordance with the results from flux compactification of IIB string theory, as an outcome of the master equation. In the present situation we interpret them as local field strengths of their respective gauge potentials and more importantly they are treated on the same footing as the respective symmetric part ($g$ or $G^{-1}$, the open and closed strings metrics), according to the open-closed strings relation. We believe that this rather uncommon perspective is equally interesting. Notice moreover that we do not resort to a doubling of the space coordinates with the coordinates corresponding to the winding modes, as in Double Field Theory \cite{Hohm:2010pp}. We are rather exploiting solely the doubling of the tangent bundle with the cotangent and the ansatz \eqref{trafo}.

\subsection{Generalized Lie bracket, connection and torsion tensor}
In this part we resume and expand the notions of generalized Lie bracket, connection and torsion tensor on sections of a vector bundle $E \xrightarrow{\pi} M$ as initially presented in our previous paper \cite{Boffo:2019zus}. The definitions for the torsion tensor and for the skew-symmetric Lie-like bracket are original. Our motivation relies on the possibility to obtain a new connection $\widetilde{\nabla} : \Gamma(E) \times \Gamma(E) \rightarrow \Gamma(E)$ from the covariant version of the Dorfman bracket, requiring that an antisymmetric bracket is subtracted from it. On the other hand, the existence of a connection whenever we are given (a symmetric bilinear form, a surjective $\rho: \Gamma(E) \mapsto \Gamma(TM)$ and) a bracket on $E$-sections which respects all the axioms of a CA bracket but Jacobi identity, and a skew-symmetric bracket (for which Jacobi identity holds instead) is also assured by the proposition enunciated in \cite{Boffo:2019zus}. The new connection $\widetilde{\nabla}$ would then be the previous one $\nabla$ summed to its torsion tensor, which is defined with the help of the Lie-like bracket. Let us begin with the introduction of an antisymmetric bracket on the sections of a vector bundle $E$.
\begin{defn} A \emph{generalized Lie bracket} (or Lie-like bracket) $\llbracket \cdot, \cdot \rrbracket$ is a bilinear operation $\llbracket \cdot , \cdot \rrbracket : \Gamma(E) \times \Gamma(E) \mapsto \Gamma(E)$ which is antisymmetric, $\llbracket U, V \rrbracket = - \llbracket V,U \rrbracket $, and respects the Leibniz rule:
\[
\llbracket U, fV \rrbracket = \rho(U) f \, V + f \llbracket U,V \rrbracket,  \quad f \in C^{\infty}(M),
\]
$\rho$ being the surjective application $\rho: \Gamma(E) \mapsto \Gamma(TM)$.
\label{lie-like}
\end{defn}
Any two generalized Lie brackets differ by an antisymmetric $2$-tensor. In analogy to the Lie brackets of vector fields, if the basis $\{\xi_{\alpha}\}$ for $E$ is holonomic, it is reasonable to set $\llbracket \xi_{\alpha}, \xi_{\beta}\rrbracket = 0$. Notice that the basis induced by the Darboux chart of the minimal symplectic realization of $E[1]$, $T^{*}[2]E[1]$, is holonomic. In this basis, being $U, V \in \Gamma(E) $, $U=U^{\alpha}(x) \xi_{\alpha} $ and $V=V^{\beta}(x) \xi_{\beta}$, a minimal example of a Lie-like bracket can be
\begin{equation}
\llbracket U,V\rrbracket = \left( \rho(U) V^{\alpha}(x) - \rho(V) U^{\alpha} \right) \xi_{\alpha}.
\label{ex-lie}
\end{equation}

The example is coordinate-dependent and thus local, but it can be made global by extending it to every patch covering the base manifold, forcing the definition \eqref{ex-lie} in the overlaps and applying the Leibniz property, that the image of $\rho$ must be a well-posed derivation, to the transition functions. Remind that the transition functions when $E$ is a $O(d,d)$-principal bundle comprise also $B$-field transformations. The whole procedure is consistent if $\rho$ is globally defined. Furthermore \eqref{ex-lie} respects Jacobi identity when $\rho$ is a homomorphism with the Lie bracket of vector fields. Our generalized Lie bracket in \cref{lie-like} is {}related to a bracket dubbed \emph{dull bracket} which was proposed in \cite{lean2012dorfman}. By definition, it fulfills the Leibniz property on both the entries and \[\rho([U,V]_{dull}) = [\rho(U),\rho(V)]_{\text{Lie}}.\] 
On the contrary, antisymmetry is not demanded. 

It would be interesting if the generalized Lie bracket could be a derived bracket. As seen in the previous section, the CA bracket is the derived bracket for the degree-$2$ symplectic structure on $T^{*}[2]T[1]M$ with homological vector field $\text{Q}$, on the algebra of degree-$1$ functions. The generalized Lie bracket \eqref{ex-lie} can be interpreted as the antisymmetrization of a derived bracket, for the canonical Poisson structure of degree $2$ equipped with an odd vector field $\text{Q}$, which fails to be Hamiltonian. For this purpose, the only non-zero Poisson brackets are:
\begin{equation}
\{p_{i}, x^{j}\} = \delta_{i}^{j} = \{\chi_{i},\theta^{j}\}.
\label{pbcan}
\end{equation}
Together with this structure we would like to consider the following $\text{\v{Q}}$:
\begin{equation}
\text{\v{Q}} := \tilde{\rho}^{\alpha i} p_{i} \dfrac{\partial}{\partial \xi_{\alpha}}.
\label{vQ}
\end{equation}
It is easy to see that $\text{\v{Q}}^{2}=0$ although it is not Hamiltonian for the symplectic form $\omega_{0}$ associated with \eqref{pbcan}: $\iota_{\text{\v{Q}}} \omega_{0} \neq \mathrm{d} \vartheta$. Then on the algebra of degree-$1$ functions the antisymmetrization of the derived bracket corresponds to the generalized Lie bracket \eqref{ex-lie}:
\begin{equation}
 \{ \text{\v{Q}} U, V\} - \{ \text{\v{Q}}V, U\} = \llbracket U, V \rrbracket .
\label{der-ll}
\end{equation}
This perspective will be helpful when we will have to determine some more or less hidden invariances pertaining the Courant algebroid.

\bigskip

The concept of a differential depending on $\rho$ is also quite important. It is first of all defined from its action on functions.
\begin{defn}
A differential $\mathrm{d}_{\rho}: \Omega^{0}(E) \cong C^{\infty}(M) \mapsto \Omega^{1}(E)$ can be given in coordinates $\{\xi_{\alpha}\}$ for the fiber of $E\xrightarrow{\pi} M$ by 
\begin{equation}
\mathrm{d}_{\rho} := \tilde{\xi}^{\alpha} \rho(\xi_{\alpha}), 
\label{drh}
\end{equation}
where $\tilde{\xi}_{\alpha}:= g_{E}^{\alpha \beta} \xi_{\beta} , \, g_{E} \equiv \langle \cdot, \cdot \rangle \in \vee^{2}E^{*}$. Then it acts on the function $f$ according to the following fully contracted formula:
\[ \langle \mathrm{d}_{\rho} f, e \rangle = \rho(e) f. \]
The action of $\mathrm{d}_{\rho}$ on the whole exterior algebra $\Omega^{\bullet}(E)$ is allowed if an antisymmetric bracket $\llbracket \cdot, \cdot \rrbracket$ as in definition \ref{lie-like} is used.
\begin{align*}
\mathrm{d}_{\rho} \psi (U_{1}, \dots , U_{k+1}) = & \sum_{i} (-)^{i} \rho(U_{i}) \psi(U_{1}, \dots , \hat{U}_{i}, \dots , U_{k+1}) \\
\, & + \sum_{i < j} (-)^{i+j} \psi(\llbracket U_{i}, U_{j} \rrbracket, \dots, \hat{U}_{i}, \dots, \hat{U}_{j}, \dots U_{k+1}).
\end{align*}
The differential $\mathrm{d}_{\rho}$ squares to zero if $\text{Im} \left(g^{-1}_{E} \circ \rho^{*} \right) = \text{Ker} \rho$ and $\llbracket \cdot, \cdot \rrbracket$ respects the Jacobi identity. 
\label{d-defn}
\end{defn}
If the existence of $\mathrm{d}_{\rho}$ is disjoint from that of $\llbracket \cdot, \cdot \rrbracket$, in the sense that a differential is given from the beginning, then it is possible to express \eqref{ex-lie} in terms of $\mathrm{d}_{\rho}$. Let $\varsigma$ be the tautological $1$-form (otherwise called soldering form) $\varsigma : E \overset{\sim}{\longrightarrow} E'$. Then the generalized Lie bracket can be obtained from:
\begin{equation} \varsigma \left( \llbracket U, V \rrbracket \right) = \rho(U) \varsigma(V) - \rho(V) \varsigma(U) - d_{\rho} \varsigma(U,V) \label{LIIE} \end{equation}
In the holonomic coordinate basis $\{\xi_{\alpha}\} = \{ \partial_{i}, \mathrm{d}x^{i}\}$ for $E = TM \oplus T^{*}M$, where $\{x^{i}\}$ are coordinates on $M$ (associated with the Darboux chart on our degree-$2$ dg-symplectic manifold), the soldering form has the expression
\[
\varsigma = \partial_{i} \otimes \mathrm{d}x^{i} \oplus \mathrm{d}x^{i} \otimes \partial_{i},
\] 
thus in \eqref{LIIE} the $\rho$-differential on $\varsigma$ is zero, and in the coordinate basis the generalized Lie bracket is \eqref{ex-lie}:
\[
\llbracket U,V \rrbracket^{i} \partial_{i} \oplus \llbracket U, V \rrbracket_{i} \mathrm{d}x^{i}  = \left( \rho(U) Y^{i} - \rho(V) X^{i} \right) \partial_{i} \oplus \left(\rho(U)\kappa_{i} - \rho(V) \sigma_{i} \right) \mathrm{d}x^{i}.
\]

\eqref{LIIE} prescribes also how the expression transforms under a change of basis.
In our companion work \cite{Boffo:2019zus} we relegated to a footnote the description of the differential $\mathrm{d}_{\rho}$, but the brackets here and there are indeed the same. 

\bigskip

At this point, with the help of a Lie-like bracket $\llbracket \cdot, \cdot \rrbracket$ many definitions of standard differential geometry for vector fields can be reproduced identically for $E$-sections too. The torsion tensor is among these.
\begin{defn}
A \emph{torsion tensor} $\text{T}: \Gamma\left(\Lambda^{2}E^{*}\right)  \rightarrow \Gamma(E)$ is defined by:
\begin{equation}
\text{T}(U,V) := \nabla_{U} V - \nabla_{V} U - \llbracket U, V \rrbracket.
\label{tors}
\end{equation} 
Because of the generalized Lie bracket \eqref{lie-like} the above defined $\text{T}$ is clearly antisymmetric and $C^{\infty}(M)$-linear.
\end{defn}

Let us now state a useful result. The proof is contained in \cite{Boffo:2019zus}.
\begin{prop}
For any bracket on $E$-sections respecting at least axioms 2 and 3 in \cref{defnCA} for a given pairing $\langle \cdot, \cdot\rangle$ and anchor map $\rho:\Gamma(E) \mapsto \Gamma(TM)$, while Jacobi identity is not required, any generalized Lie commutator as in \cref{lie-like} and any affine connection on generalized vector fields, metric with respect to the pairing, and with fully antisymmetric torsion $\text{T} \in \Gamma\left(\Lambda^{3} E^{*}\right)$ \eqref{tors}, the following relation holds:
\begin{equation}
\langle [U,V] - \llbracket U, V \rrbracket , W \rangle = \langle \nabla_{W} U, V \rangle.
\label{THM}
\end{equation}
Notice that ultimately the above relation says that the Dorfman bracket for a CA $[\cdot,\cdot]_{\text{D}}$, a generalized Lie bracket $\llbracket \cdot, \cdot \rrbracket$ and a connection $\nabla$ are dependent on each other. \label{pr}
\end{prop}

We will be interested exclusively in extracting a covariant derivative from given $[\cdot,\cdot]$ and $\llbracket \cdot, \cdot \rrbracket$, using the proposition. It might hence be helpful to go through the evidences that the derivative operator $\nabla$ \eqref{THM} has indeed the correct properties:
\begin{equation}
\begin{matrix*}[l]
\left\langle \nabla_{fW} U , V \right\rangle = \left\langle \nabla_{W} U, fV \right\rangle = f \left\langle \nabla_{W} U, V \right\rangle, & \left\langle \nabla_{W} fU, V \right\rangle = \rho(W) f \langle U, V\rangle + f \left\langle \nabla_{W} U, V \right\rangle,\\
\rho(W) \langle U,V \rangle = \langle \nabla_{W} U, V\rangle + \langle U, \nabla_{W} V\rangle, & \langle \text{T}(U,V), V\rangle =0.
\end{matrix*}
\label{def_con}
\end{equation}
Tensoriality in $W$ is obvious. Tensoriality in $V$ is due to the fact that $[\cdot,\cdot]$ and $\llbracket \cdot,\cdot \rrbracket$ respect the Leibniz property in their respective second slot. $\mathbb{R}$-linearity in $U$ is due to:
\[
[fU,V] = -\rho(V)f \, U + f [U,V] + \mathrm{d}_{\rho}f \, \langle U,V\rangle,
\]
which stems from the axiom about the symmetric part of the bracket and Leibniz rule for $\rho$. Then, metricity of the connection is an outcome of the third axiom for a CA bracket together with the antisymmetry of the generalized Lie bracket. The torsion is completely skew-symmetric due to both the conditions (2 and 3) for the bracket $[\cdot,\cdot]$ in \cref{defnCA}, and because $\llbracket \cdot,\cdot \rrbracket$ is antisymmetric. 

Now we want to employ some of the brackets previously discussed in the proposition to straightforwardly yield a new connection $\widetilde{\nabla}$. As for the bracket respecting two of the CA axioms we pick up the deformed Dorfman bracket \eqref{comp_2}. Recall that it comes from the derived Poisson brackets with undeformed Hamiltonian $\Theta$. Then we are led to express the Lie-like bracket w.r.t.\,the same holonomic coordinate basis, while employing the new anchor. Hence $\llbracket \cdot, \cdot \rrbracket$ in \eqref{ex-lie} does the job. With these elements, the connection $\widetilde{\nabla} : \Gamma(E) \rightarrow \Gamma(E^{*}) \otimes \Gamma(E)$ corresponds to
\begin{equation}
\left\langle \widetilde{\nabla}_{W} U, V \right\rangle^{\prime} = \left\langle \text{T}(U,V) , W \right\rangle^{\prime} + \left\langle \nabla_{\rho(W)} U, V \right\rangle^{\prime}.
\label{CONNECT}
\end{equation}
As displayed here, the torsion tensor for the connection $\nabla$ we started with, emerges from the fully general contracted expression of the Dorfman bracket \eqref{DORF} upon subtraction of $\langle \llbracket U, V \rrbracket , W \rangle^{\prime}$. We now have a perfectly consistent Courant algebroid connection on $E$. However for our scope the relevant connections are obtained by restriction of the ``big'' connection to tangent or cotangent space exclusively. In both the resulting connections the direction of derivation will be anchored to tangent space by means of the anchor. 

\subsection{Connections on tangent space and on cotangent space for the given deformation}

In this part we are going to halve the dimension of the fiber of our bundle by restricting our considerations to $TM$ or $T^{*}M$ respectively. The case of $TM$, analyzed extensively in many other works e.g.\,\cite{Jurco:2015ywk}, is presented just for reference and is regarded as dual to the case of $T^{*}M$, which is the relevant one for the description of a gravitational field paired to the local expressions of the stringy non-geometric fluxes. As subspaces of $E$, $TM$ and $T^{*}M$ are isotropic with respect to the natural pairing $\eta$ of vector fields with  forms but not with respect to the metric $\mathcal{G} \equiv \langle \cdot , \cdot \rangle'$. They are involutive under the standard Dorfman bracket but not under the non-canonical one. 

A clever way to project onto tangent or cotangent space utilizes a splitting $s: \Gamma(TM) \rightarrow \Gamma(E)$, $\rho \circ s = \text{id}_{TM}$, or an embedding $r: \Gamma(T^{*}M) \rightarrow \Gamma(E)$. $r$ itself can be considered a splitting of the dual short exact sequence
\[
0 \rightarrow TM \xrightarrow{\Delta} E \xrightarrow{\Delta^{*}} T^{*}M \rightarrow 0,
\]
if it fulfills the condition $\Delta^{*} \circ r = \text{id}_{T^{*}M}$. If $\rho$ is given by \eqref{dual-anchor}, a consistent choice for $\Delta^{*}$ can be\footnote{We are requiring that $\rho \circ \Delta = \text{id}$.} 
\begin{equation}
\Delta^{*}(U) = -\left(g-B\right)(X) + \sigma.
\end{equation} 
Of all the possible splittings, we are going to use only the trivial embeddings, $s: TM \mapsto E$: 
\begin{equation}
s(X) = \begin{pmatrix}X \\ 0 \end{pmatrix} \in E \cong TM \oplus T^{*}M; \label{non-iso-s}
\end{equation} 
and correspondingly, $r: T^{*}M \mapsto E$:
\begin{equation}
r(\sigma) = \begin{pmatrix} 0 \\ \sigma \end{pmatrix} \in E \cong TM \oplus T^{*}M. \label{r-splitting}
\end{equation}
As announced, the induced metrics on $TM$ and $T^{*}M$ are non-degenerate and corresponds to:
\[
\langle s(X), s(Y) \rangle = -2g(x,Y), \quad \langle r(\sigma), r(\zeta)\rangle = 2 G^{-1}(\sigma, \zeta).
\]
Eventually we can force our connection \eqref{CONNECT} on tangent (cotangent) space by \emph{defining} it in the following way:
\[
\langle \widetilde{\nabla}_{s(Z)} s(X), s(Y) \rangle' =: -2g\left( \widetilde{\nabla}_{Z} X, Y \right), \quad \langle \widetilde{\nabla}_{r(\zeta)} r(\sigma) , r(\kappa) \rangle'=: 2G^{-1} \left( \widetilde{\nabla}_{\zeta} \sigma, \kappa \right).
\]
Let us now briefly comment upon the regular connection on vector fields first. Our list of prescriptions (the new Dorfman bracket homomorphic to the canonical one via $\mathcal{E}$, the anchor $\rho$, the generalized Lie bracket and the splitting $s$ which leads to a non-degenerate induced metric on $TM$) yield:
\begin{equation}
2g\left( \widetilde{\nabla}_{Z} X , Y \right) = \left(g+B\right)\left([X,Y],Z\right) + \iota_{Z} \left( \mathcal{L}_{X} \left(g-B\right)(Y) - \iota_{Y}d\left(g-B\right)(X)\right) -2g([X,Y],Z) , \, \label{vecs}
\end{equation}
where the last term is due to the generalized Lie bracket (fully contracted to a $C^{\infty}(M)$-function). It is hence immediate to show that the connection for tangent space is the Levi-Civita one (metric w.r.t.\, $g$ and torsion-free) plus an antisymmetric term based on the exterior derivative of $B$, the $3$-form $H = \mathrm{d} B$:
\begin{equation}
\widetilde{\nabla}_{Z} X = \nabla^{\text{L.C.}}_{Z} X - \dfrac{g}{2}^{-1}  \circ H(Z,X, \cdot),
\label{on_vf}
\end{equation}
Its torsion is the fully antisymmetric $H$ itself. The reader should be aware that thanks to our systematic method the connection is a globally defined object, although the starting point (graded symplectic structure with a homological vector field) was not.

We can now turn our attention to cotangent space, and study this instance of derived connection which, following reference \cite{Asakawa:2015jza}, we would like to call ``contravariant'' connection. Let us first remind the reader that the anchor $\rho$ provides a natural derivative operator in the direction of forms,
\[
\rho(r(\sigma)) = \left( G^{-1} + \Pi \right)(\sigma) : C^{\infty}(M) \rightarrow \Gamma( TM ) .
\]
Hence using the splitting $r$ of the dual sequence we obtain the connection $\widetilde{\nabla}: \Gamma(T^{*}M) \mapsto \Gamma(TM) \otimes \Gamma(T^{*}M)$:
\begin{align}
2 G^{-1} \left( \widetilde{\nabla}_{\zeta} \sigma, \kappa \right) =&  \iota_{\left[ \rho(r(\sigma)), \rho(r(\kappa)) \right] } \zeta + \iota_{\rho(r(\zeta))} \left( \mathcal{L}_{\rho(r(\sigma))} \kappa - \iota_{\rho(r(\kappa))} d \sigma \right)\notag \\
& - 2G^{-1}\left( \rho(r(\sigma)) r(\kappa) - \rho(r(\kappa)) r(\sigma), r(\zeta) \right),  \label{some-name}
\end{align}
where the last row is due to the generalized Lie bracket. The expression might look unwieldy, however the right properties of an affine connection \eqref{def_con} are checked. Notice that the direction of derivation, given by $\zeta$, is always anchored to $TM$, via $\rho \circ r: \Gamma(T^{*}M) \mapsto \Gamma(TM)$. In the generalized Lie bracket this happens thanks to an easy manipulation
\[
2G^{-1}\left( \dots , r(\zeta) \right) = 2g\left(G^{-1} + \Pi\right) \left(\dots , \rho(r(\zeta)) \right),
\]
while the first addend in \eqref{some-name} can be massaged to
\[
\iota_{\left[ \rho(r(\sigma)), \rho(r(\kappa)) \right] } \zeta \equiv 2G^{-1} \left( \dfrac{G}{2} \left( \left[ \rho(r(\sigma)), \rho(r(\kappa)) \right] \right), \zeta \right) = G \left( \left[ \rho(r(\sigma)), \rho(r(\kappa)) \right] , G^{-1}(\zeta) \right).
\]
In the coordinate basis the connection becomes definitely more familiar. Assigning indices $i, j, k$ to $\sigma, \kappa $ and $\zeta$ respectively, and using $G^{-1}$ to raise or lower indices,
\begin{equation}
\mathit{\tilde{\Gamma}}^{ki}{}_{j} = \dfrac{1}{2}G_{jl} \left( 2 \left(G^{-1} + \Pi \right)^{[i \vert m} \partial_{m} \left(G^{-1} + \Pi \right)^{\vert l ] k} + \left(G^{-1} + \Pi \right)^{km} \partial_{m} \left( G^{-1} + \Pi\right)^{il} \right).
\label{on_forms}
\end{equation}
The expression in components now involves the $R$\footnote{$R$ is as well $R \equiv 3[\Pi, \Pi]_{\text{S}}$, where $[,]_{\text{S}}$ is the Schouten bracket of multivector fields.} and $Q$ fluxes, which according to \cite{Andriot:2012wx} in the supergravity frame are 
\begin{align}
R^{ijk} := & 3 \Pi^{[i \vert l} \partial_{l} \Pi^{\vert jk]} , \label{RR}  \\ Q^{ij}{}_{k} := & \partial_{k} \Pi^{ij}.
\end{align}
The connection coefficient \eqref{on_forms} consists of a symmetric part, equivalent to the Christoffel symbol $\Gamma_{G}$ for $G^{-1}$ and to another triplet of terms. We will dub $\mathcal{Y}$ the following:
\[
\mathcal{Y}^{ki}{}_{j} := \frac{1}{2} G_{jl} \left( \Pi^{km} \partial_{m} G^{il} + 2\Pi^{[i \vert m} \partial_{m} G^{\vert l] k} \right), \quad \mathcal{Y}^{ki}{}_{j} = \mathcal{Y}^{ik}{}_{j} , \quad \mathcal{Y}^{ki}{}_{k} = \frac{1}{2} G_{lm} \Pi^{ip} \partial_{p} G^{lm} .
\]
Together, $\Gamma_{G}$ and $\mathcal{Y}$ build up a metric and torsionless connection for the metric $2G^{-1}$. It can be convenient to rewrite them so that the anchor is prominent:
\[
\Gamma_{G}^{ki}{}_{j} + \mathcal{Y}^{ki}{}_{j} = \dfrac{1}{2} G_{mj} \left(\rho^{k l} \partial_{l} G^{im} + \rho^{il} \partial_{l} G^{mk} - \rho^{jl} \partial_{l} G^{ik}\right) .
\]
The torsion of the connection is made up by derivatives hitting $\Pi$, that lead to the $R$ tensor and to a triplet of $Q$ terms, which for later convenience we denote in the following notation:
\begin{equation}
\mathcal{X}^{ki}{}_{j} := -\frac{1}{2} Q^{ki}{}_{j} +  G_{jl} G^{[k \vert  m} Q^{ \vert i ]l}{}_{m}, \quad \mathcal{X}^{ki}{}_{j} = - \mathcal{X}^{ik}{}_{j}, \quad \mathcal{X}^{ki}{}_{k} = 0. \label{XX}
\end{equation}
Later, when needed, we will untie $\mathcal{X}$ in its $Q \sim \partial \Pi$ constituents. Eventually the connection symbols \eqref{on_forms} are given by the combination
\begin{equation}
\mathit{\tilde{\Gamma}}^{ki}{}_{j} =  \Gamma_{G \; \; \; j}^{\; \; \; ki} + \mathcal{Y}^{ki}{}_{j} + \frac{1}{2} G_{jl} R^{kil} +\mathcal{X}^{ki}{}_{j}  .
\label{Form_Con}
\end{equation}
The torsion tensor of $\widetilde{\nabla}$ is
\[ \text{T}^{ijk} = 2 R^{ijk} + 4 \mathcal{X}^{ijk}, \]
and it is easy to see that it is fully antisymmetric, as predicted in proposition \eqref{pr}.

\subsection{Curvature tensor}
In this part we would like to present our working definitions for the analogue of the Riemann tensor and of Ricci tensor, specialized to the relevant subbundles for this article. First we will sketch how to get a genuine Riemann curvature tensor for the $E$-connection $\widetilde{\nabla}$ on the generalized tangent bundle $E$. Usually in Generalized Geometry (see for example \cite{Coimbra:2011nw}) a curvature tensor, built with the non-skewsymmetric bracket for the CA, is defined in the special circumstance that the generalized metric $\mathcal{H}$ is introduced, and its expression needs to be evaluated on a combination of the subbundles associated with $\mathcal{H}$. Here instead it can mimic the Riemann tensor of standard differential geometry and be simply the commutator of the covariant derivative $\widetilde{\nabla}$ \eqref{CONNECT} minus the covariant derivative of the commutator, where the latter is constructed with $\llbracket \cdot, \cdot \rrbracket$.
$\text{Riem} \in \Gamma(\overset{3}{\bigotimes} E^{*} \otimes E)$ is thus
\begin{equation}
 \left[\widetilde{\nabla}_{U}, \widetilde{\nabla}_{V} \right] W - \widetilde{\nabla}_{\llbracket U,V\rrbracket} W =: \text{Riem}(U,V,W) \in \Gamma(E).
\end{equation}
The definition is well-posed, i.e.\,$\text{Riem}$ is a true rank-$4$ tensor in $E$. This claim is easily checked from the behaviour under function multiplication of the commutator of covariant derivatives and of the generalized Lie bracket. This definition is definitely much more straightforward than the most common one illustrated before. 

In the end we are interested in specializing $\text{Riem}$ in the obvious way to the regular connection on tangent vectors only, and to that on covectors. Their curvatures stem from evaluating $\text{Riem}$ with $s$-vectors only or $r$-forms only, and project the result on the corresponding subbundle. For example, in the $TM$ case the generalized Lie bracket when the splitting $s$ is given by \eqref{non-iso-s}, matches the Lie bracket of vector fields
\begin{equation}
 \text{Riem}_{TM}(X,Y,Z) := \rho\left( \text{Riem} (s(X),s(Y), s(Z)) \right) \overset{\eqref{non-iso-s}}{=} \left[\widetilde{\nabla}_{X}, \widetilde{\nabla}_{Y} \right] Z - \widetilde{\nabla}_{[X,Y]} Z .
\end{equation}
The Riemann curvature tensor $\text{Riem}_{T^{*}M} \in \Gamma(\otimes^{3}TM \otimes T^{*}M)$ corresponds to:
\begin{equation}
\text{Riem}_{T^{*}M}(\zeta, \kappa, \sigma) := \Delta^{*}\left(\text{Riem}(r(\zeta), r(\kappa), r(\sigma))\right)  \overset{\eqref{r-splitting}}{=} \left[ \widetilde{\nabla}_{\zeta}, \widetilde{\nabla}_{\kappa} \right] \sigma - \widetilde{\nabla}_{\llbracket r(\zeta), r(\kappa)\rrbracket} \sigma. \label{RIC}
\end{equation}
The Ricci tensor $\text{Ric}_{TM} \in \Gamma(T^{*}M \otimes T^{*}M)$ (resp.\,$\text{Ric}_{T^{*}M} \in \Gamma(TM \otimes TM)$) are obtained from the partial trace of $\text{Riem}_{TM}$ (resp.\,$\text{Riem}_{T^{*}M}$):
\[
\text{Ric}_{TM}(Y,Z) = \sum_{i}^{d} \langle \text{Riem}_{TM}\left(\partial_{i}, Y,Z \right) , \mathrm{d}x^{i}\rangle, \quad \langle \cdot, \cdot\rangle \; \text{canonical pairing.}
\]
In the next section we can present a pair of action functionals for a peculiar contraction of $\text{Ric}_{TM}$ and of $\text{Ric}_{T^{*}M}$.

\section{Gravitational actions with fluxes}
\label{E}

This section addresses the application to physics of the algebraic and geometric methods developed in the previous sections. Before showing and discussing the gravitational action involving $G^{-1}$ and the $Q$ and $R$-fluxes, let us briefly present the result for the connection on tangent space \eqref{on_vf}. The Ricci tensor of that connection, integrated against $g^{-1} \circ \left(g-B\right) \circ g^{-1}$, and by means of Stokes' theorem in the integration, is known to reproduce the effective action functional of the low-energy closed strings, i.e.\;the Ricci scalar and the square of the field strength $H$ \cite{Jurco:2015ywk}, \cite{Boffo:2019zus},
\begin{equation}
S_{NS}[g, B] = \int_{M} \text{vol}_{g} \, R(g) - \dfrac{1}{12} H^{2}.
\label{S-NSN}
\end{equation}
The above formula is the Hilbert-Einstein action of the connection \eqref{on_vf}, formed with the non-symmetric combination $g-B$. It could be derived deploying elements of differential geometry of $TM \oplus T^{*}M$, as well as the $\mathcal{E}(x)$-deformed graded Poisson algebra with its Q structure.

A similar result can now be unveiled for the contravariant connection \eqref{on_forms}. It concerns a gravitational field together with a kinetic term for the field strength of $\Pi$, where the exterior derivative is taken with the anchor and hence on the $1$-forms corresponds to $R$ and $\mathcal{X}$ as in our formulas \eqref{RR} and \eqref{XX}. These are the local expressions of the stringy $R$ and $Q$ fluxes. The invariant Lagrangian is given by a scalar built from the Ricci curvature for the $T^{*}M$-subbundle. In components the Ricci tensor $\text{Ric}_{T^{*}M} \in \Gamma(TM \otimes TM)$ from \eqref{RIC} is
\begin{equation}
\text{Ric}^{jk}_{T^{*}M} = \left(G^{-1} + \Pi \right)^{[l\vert m} \partial_{m} \mathit{\tilde{\Gamma}}^{\vert j]k}{}_{l} + \mathit{\tilde{\Gamma}}^{[l \vert m}{}_{l} \mathit{\tilde{\Gamma}}^{ \vert j] k}{}_{m},
\end{equation}
where indices are contracted by means of the identity matrix of $\text{End}(T^{*}M)$. Notice also that the antisymmetrization is performed here without factors of $\frac{1}{2}$. Let us abbreviate the symmetric part of the connection with the covariant derivative $\tilde{D}_{G}$: 
\[
\tilde{D}_{G}^{i} \varsigma := \left(G^{-1} + \Pi \right)^{im} \partial_{m} \varsigma + \varsigma_{k} \left(\Gamma_{G\;\; \; j}^{\; \; \; ik} + \mathcal{Y}^{ik}{}_{j} \right) dx^{j}.
\]
Then the Ricci tensor is:
\begin{equation}
\text{Ric}^{jk}_{T^{*}M} =  \, \text{Ric}_{G}^{jk} - \frac{1}{4} R^{jm}{}_{l} R^{lk}{}_{m}   - \frac{1}{2} R^{(j \vert m}{}_{i} \mathcal{X}^{i \vert k)}{}_{m}  - \mathcal{X}^{ j m}{}_{i} \mathcal{X}^{ ik}{}_{m}   + \frac{1}{2} \tilde{D}_{G}^{i} R^{jk}{}_{i} + \tilde{D}_{G}^{i} \mathcal{X}^{jk}{}_{i} .
\label{beta}
\end{equation}
In this expression, the first four terms are completely symmetric, while the last two are purely antisymmetric. $\text{Ric}_{G}$ is the Ricci tensor, on the cotangent bundle, for the symmetric part of the connection, i.e. it is the commutator of a pair of $\tilde{D}_{G}$. It is worth to stress again that it is the curvature tensor for a connection on forms, metric with respect to $G^{-1}$. The explicit expression for $\text{Ric}_{G}^{jk}$ is:
\[
\text{Ric}_{G}^{jk} = \left( G^{-1} + \Pi \right)^{[l \vert m} \partial_{m} \left( \Gamma_{G \; \; \; l}^{\; \; \vert j] k} + \mathcal{Y}^{ \vert j] k}{}_{l} \right) + \left( \Gamma_{G \; \; \; \; \; \; l}^{\; \;  [l \vert m} + \mathcal{Y}^{ [l \vert m}{}_{l} \right) \left( \Gamma_{G \; \; \; \, m}^{\; \; \vert j] k} + \mathcal{Y}^{\vert j] k}{}_{m} \right).
\]
To retain the antisymmetric terms of the Ricci curvature tensor we can contract $\text{Ric}_{T^{*}M}$ with the volume form and the non-symmetric $\left(G^{-1} - \Pi\right)$, in this combination:
\begin{equation}
\sqrt{ \det G^{-1} }  \left(G \circ \left( G^{-1} - \Pi \right) \circ G \right)_{jk} .
\label{contraction}
\end{equation}
In fact the following relation holds:
\[
 \left( G^{-1} + \Pi \right)^{jl} \partial_{l} \left( \sqrt{\det G^{-1}} \, w \right)  = \sqrt{  \det G^{-1}} \, \left( \Gamma_{G \; \; \; k}^{\; \; \; kj} + \mathcal{Y}_{\; \; \; k}^{kj} \right) w= \sqrt{\det G^{-1}} \, \tilde{D}_{G}^{j}w \equiv \sqrt{\det G^{-1}} \, \widetilde{\nabla}^{j} w.
\]
This says that the total derivative of the volume form times a scalar is equal to the divergence, accompanied with the scalar half-density $\sqrt{\det G^{-1}}$. Hence we can perform integration by parts on the contracted action, use Poincar\'{e} lemma for the boundary term and implement the following observation:
\[
\tilde{D}^{l}_{G} \Pi_{jk} \equiv G_{jn} G_{rk} \left( G^{-1} + \Pi \right)^{lm} \partial_{m} \Pi^{nr} = G_{jn} \left(\frac{1}{3} R^{lnr} + G^{l m} Q^{nr}{}_{m} \right) G_{rk}.
\]
Eventually we end up with the action functional\footnote{Despite the possible confusion, we deploy $\mathrm{d}x$ to allude to a $d$-dimensional top form.}:
\begin{align}
S[G^{-1},\Pi] = \int_{M} \mathrm{d}x \,  \sqrt{\det G^{-1}}  & \bigg[ \text{R}_{G} - \frac{1}{12} R^{2} + \dfrac{3}{4} \left( Q^{jn}{}_{m}  \left( G^{pm} Q^{lk}{}_{p} - 2 Q^{ml}{}_{p} G^{pk} \right) \right) G_{jl} G_{nk} \notag \\
 \, &  - \left(\dfrac{1}{6} Q^{jn}{}_{s} + \dfrac{2}{3} G_{rs}G^{j p} Q^{nr}{}_{p} \right)R^{slk}  G_{jl} G_{nk} \bigg] . \label{action}
\end{align}
It comprehends therefore the relevant tensors that are T-dual to the $H$-flux in compactifications (e.g.\,on a $\mathbb{T}_{3}$ torus) of string theory (NS-NS sector, with zero dilaton). It is manifestly invariant under diffeomorphisms. The action \eqref{action} is the Einstein-Hilbert term for a connection living in the cotangent space solely, and built with the open string metric and bivector $\Pi$. It is obtained through the same mechanism with which we managed to rebuild the NS-NS sector of closed string effective action for zero dilaton \eqref{S-NSN}, from the Levi-Civita connection with $H$ torsion $\nabla^{\text{L.C.}} -\frac{1}{2} g^{-1}\circ H$. Also, the final contraction retained the antisymmetric part as well. 

\subsection{Invariance under gauge symmetries of $\Pi$}

The gravitational actions that we constructed are manifestly invariant under diffeomorphisms, but this is not the end the story. The first action \eqref{S-NSN} is also invariant under the gauge symmetry of the $B$ field, $B \mapsto B + \mathrm{d} \lambda$. Therefore one could imagine that the second action \eqref{action} might be invariant under the gauge symmetry of $\Pi$.

As anticipated in section \ref{g-syms}, the generators of gauge symmetries are $\text{Q}$-exact degree $2$ functions $\alpha$ quadratic in the $\xi$ coordinates. Though for the non-symmetric theory of gravity \eqref{S-NSN} we can guess them easily by knowing that $H = \mathrm{d} B$, for $S[G^{-1},\Pi]$ \eqref{action} we must crank the machinery of \cref{g-syms}, and find $\alpha$ from \eqref{qex} subjected to the condition \eqref{var0}. The kernel of our anchor map comprises the sections:
\begin{equation}
\Upsilon := \begin{pmatrix} \varrho \\ - \left(G^{-1}+\Pi\right)^{-1}(\varrho) \end{pmatrix}, \quad \varrho \in \Gamma(TM).
\label{kha}
\end{equation}
These sections are the graph of $-\left(G^{-1} + \Pi\right)^{-1} = - (g+B)$. However, also the graph of the inverse map could be taken to characterize the kernel of the anchor. The result is independent of the choice, but the computation can gain some feasibility. Expression \eqref{kha} is very efficient to derive the gauge symmetries of $B \in \Omega^{2}(M)$ in the graded geometric setting, where they can be easily recognized as the exterior derivative of a $1$-form. To better tackle the case of $\Pi$, we can resort to the following expression for the sections in the kernel of $\rho$: 
\begin{equation}
\Upsilon := \begin{pmatrix} -\left(G^{-1} +\Pi\right)(\varsigma) \\ \varsigma \end{pmatrix}, \quad \varsigma \in \Gamma(T^{*}M).
\end{equation}
Now we must carefully select the $\text{Q}$-exact sections $\alpha$,
\[
\alpha^{\beta \gamma} = \tilde{\rho}^{\beta i} \nabla_{i} \Upsilon^{\gamma},
\] 
that belong to the subspace of quadratic functions in $\theta$. In fact in the deformed Poisson algebra, dualization occurs by means of $G^{-1}$. Using the connection symbols in the second row of \eqref{matrix-conn} and sending $\theta$ into $G^{-1} \chi$ we end up with the sought-after $\delta \Pi^{nm} = \alpha^{nm}$:
\begin{align}
\delta\Pi^{jk} = \dfrac{1}{2} \left(G^{-1}+\Pi\right)^{jl} \bigg(& \varsigma_{i} \partial_{l} \left(G^{-1}+\Pi\right)^{ik} - \left(G^{-1} + \Pi\right)(\varsigma)^{i} \left(G^{-1} - \Pi\right)^{-1}_{im} \partial_{l} \left(G^{-1} - \Pi\right)^{mk} \notag \\
\, & +2G^{km} \partial_{l} \varsigma_{m}(x) \bigg).
 \label{deltaP}
\end{align}
The gauge symmetry of $\Pi$ \eqref{deltaP} looks a bit contrived, but it would be plain wrong to guess $\Pi \mapsto \Pi + \mathrm{d}_{\rho} \varsigma$, because the bivector sources both the $Q$-flux and the $R$-flux in our model.

We have thus represented the gauge symmetry of $\Pi$ \eqref{deltaP} in the graded symplectic setting. But then the Dorfman bracket, being the derived bracket of the graded Poisson bracket, preserves this very same symmetry. Now to be able to conclude that the whole construction, which relies on proposition \ref{pr}, does not spoil the symmetry, we must verify that the generalized Lie bracket is invariant under the same symmetry of the Dorfman bracket. For this to happen, the derived bracket construction is again useful: according to the derived bracket formula for the generalized Lie bracket \eqref{der-ll}, $\varsigma_{i}(x) \left( -\left(G^{-1} +\Pi\right)^{ij} \chi_{j} + \theta^{i} \right)$ must be $\text{\v{Q}}$-closed \eqref{vQ}. With our anchor map, $\text{\v{Q}}$ is:
\[
\text{\v{Q}} = p_{i} \dfrac{\partial}{\partial \chi_{i}} + \left(G^{-1} + \Pi\right)^{ji} p_{i} \dfrac{\partial}{\partial \theta^{j}}.
\]
The conclusion is immediate: $\Upsilon$ is closed under the homological vector field $\text{\v{Q}}$ which fails to be Hamiltonian. Hence, requiring that also every $\Pi$ involved with the anchor map is left unchanged, $R$ \eqref{RR} and $\mathcal{X}$ \eqref{XX} are invariant under the transformation $\Pi \mapsto \Pi + \delta \Pi$. Invariance of the action \eqref{action} follows. 

A similar action with non-geometric fluxes, that is also invariant under the gauge symmmetry of the bivector $\Pi$, can be found in \cite{Andriot:2013xca} and \cite{Blumenhagen:2013aia}. Under different conditions, the work \cite{Asakawa:2015jza} proposed a gravitational theory for a metric in the background of the $R$-flux, exhibiting invariance under the gauge symmetries of a Poisson bivector. The authors used a connection on a Poisson Courant algebroid (i.e.\;a CA twisted with a Poisson bivector). Its curvature scalar consists of the Ricci scalar for what in our notation is $\Gamma_{G}$ and of the square of a $R$ flux given by a Poisson bivector $\theta$ and our $\Pi$ as $R := [\theta, \Pi]_{\text{S}}$. They proved the invariance of the gravity action under $\Pi \mapsto \Pi + \mathrm{d}_{\theta} \pi$, where the differential is the same as the $\rho$-differential used here and $\theta$ serves as an anchor map.

\section{Discussion and outlook}
\label{C}
Let us conclude the article with some comments on the achievements, some remarks on the important details and an outlook on future research.

The purpose of the work was twofold: On one hand, Moser lemma allowed us to select a non-canonical graded symplectic structure, involving the geometric data of a metric and a $2$-form on the space of degree-$1$ (Grassmann) variables. The other coordinates of the graded manifold were not transformed. The degree-$1$ objects therefore ceased to be $O(d,d)$-vectors but they rather belonged to a $O(d) \times O(d)$ representation. On the other hand, the derived bracket of the Poisson brackets with Hamiltonian $\Theta = \xi_{\alpha} \tilde{\rho}^{\alpha i} p_{i}$, $\tilde{\rho}^{\alpha i}$ in \eqref{anch-def} and $C^{\alpha \beta \gamma} = 0$, alas the Dorfman bracket of the exact CA $E \cong TM \oplus T^{*}M$, had a simple coordinate expression in which covariant derivatives replaced the partial ones. The connection giving rise to a covariant derivative is due to the modified Poisson bracket between degree $1$ functions and degree $2$ linear functions of $p$ (see \eqref{def-expl}). The CA bracket is a derivation on sections, but when the antisymmetric Lie-like bracket on the generalized vector fields \eqref{ex-lie} is subtracted it becomes an affine connection $\widetilde{\nabla}$. This new connection turns out to be given by the previous covariant derivative added to its torsion tensor (defined with the antisymmetric $\llbracket \cdot, \cdot \rrbracket$). We studied $\widetilde{\nabla}$ in two ``limits'': when all of its arguments were projected onto $TM$, and when they were projected onto $T^{*}M$. Since the induced metrics on $TM$ and $T^{*}M$ are non-degenerate, we obtained the connections $\widetilde{\nabla}: \Gamma(TM) \mapsto \Gamma(T^{*}M \otimes TM)$ \eqref{on_vf} and $\widetilde{\nabla}: \Gamma(T^{*}M) \mapsto \Gamma\left(TM \otimes T^{*}M\right)$ \eqref{on_forms}. The curvature invariants for these were shown to {}yield Lagrangians of physical interest. In the former case, we {}have recovered the NS-NS sector of low-energy closed string theory (without dilaton), from a curvature scalar that retained the antisymmetric contribution. Something similar happened in the latter case, upon integrating the Ricci curvature against a non-symmetric tensor. The action obtained in that way consisted of the Ricci scalar for the Levi-Civita connection of the contravariant metric $G^{-1}$ (where however the anchor comprises both $G^{-1}$ and $\Pi$) and other invariant terms built with the local expressions of the $Q$ and $R$ fields of string compactification. Moreover $S[g,B]$ \eqref{S-NSN} and $S\left[G^{-1}, \Pi\right]$ \eqref{action} hold for an arbitrary number of dimensions.

The findings presented here share some similarities with other works on the subjects, and they differ in some aspects.
We focused on a very minimal deformation of the graded symplectic structure, as only the algebra of degree $1$ functions was reshaped. In this way, and by the chosen particular deformation with a $2$-form $B$ and a bivector $\Pi$, dependent on $B$, we could interpret the stringy 3-form $H$ and the connection coefficients $Q$ and $R$ (where the first has 1 contravariant and 2 covariant indices, and the latter is a trivector) as formal ``field strengths'' of $B$ and $\Pi$. In our companion work \cite{Boffo:2020vqx} we considered a deformation based only the open string quantities $G^{-1}$ and $\Pi$, and obtained the same action functional. Usually in dg-symplectic geometry the stringy fluxes appear in the Hamiltonian, and are responsible for a twisting of the Dorfman bracket of the associated CA. Our approach via deformation yields global objects (connections, actions) and does not leave the realm of (more or less) standard geometry in favor of what is conventionally known as non-geometry. Instead this happens for backgrounds with $R$-flux: local patchings cannot be glued together. Instead our $R$ is merely the {}dual exterior derivative of the bivector $\Pi$ and therefore a standard trivector, and that explains why it was possible to suggest an action for a contravariant symmetric tensor in the presence of $\Pi$. 

Deforming the $2$-graded Poisson algebra gives a technique to include interactions with gauge potentials, which is most commonly deployed for the phase space algebra of a point particle in a given electromagnetic field. Locally the (graded) Poisson algebra has the facets of a gauge theory, i.e.\,a principal bundle with some structure group, and in this perspective some problematic aspects can be tackled in an easier manner: For example, the new graded Poisson algebra \eqref{PB-new} can be quantized. Its quantization can be of Weyl type or Clifford type, depending on the parity of the algebra of functions. Anyway this quantization concerns the particle in a non-flat background having interactions with $B$ and $\Pi$, not the gravity theories themselves.
 
Our construction differs from the T-dual field redefinition of the NS-NS bosonic closed string action \cite{Andriot:2013xca}, its Lie algebroid description \cite{Blumenhagen:2013aia}, \cite{Blumenhagen:2012nk} and their respective generalized Ricci scalars, for example because of our different choice of the basis of the fiber, that treats equally the metric and the antisymmetric field, and also because of the introduction of the generalized Lie bracket. Another big difference is due to the simultaneous presence in the ansatz for the vielbein of $g-B$ and $\left(g+B\right)^{-1}$, so that the anchor map is forced to be $(G^{-1} + \Pi)(\sigma)$, while so far just $\Pi(\sigma)$ was employed as anchor in the above references. Clearly if we had to choose just $\iota_{e^{a}}\Pi$ as new basis for the $1$-forms we would not get a block diagonal generalized metric on the CA $E$. Notice that our systematic method returns a well-defined connection on generalized vectors, provided that a generalized Lie bracket is given, and it is simple to define a curvature tensor in our framework. This is not so obvious in Generalized Geometry. We did not double the dimension of the base manifold. A further analysis on the relations with DFT is beyond the scopes of this work: however it is rather reasonable to expect our action functional to be some limit of the DFT invariant action of Hohm, Hull and Zwiebach \cite{Hohm:2010pp} when the section condition is applied. A similar setting, namely a graded Poisson algebra with the trivector $R$, was related to the section condition of DFT in \cite{Heller:2017mwz}.

We currently lack a construction that takes the R-R fields into account. We nevertheless expect these fields, which are tensors of the spin bundle, to be treatable within the realm of a graded Poisson structure of degree $2$; they have been already described in Generalized Geometry for example in \cite{Severa:2018pag} and \cite{Coimbra:2011nw}.\\

\smallskip

\textbf{Acknowledgements}: E.B. and P.S. are grateful to the RTG 1620 ``Models of Gravity'' for funding. E.B. also thanks GA\v{C}R Grant EXPRO 19-28628X for support. The authors want to thank A. Chatzistavrakidis and M. Pinkwart-Walker for discussion and suggestions on the work. Thanks to M. Jotz Lean for bringing the dull bracket to our attention.

\bibliographystyle{ieeetr}
\bibliography{biblio_T-dual_action}

\begin{thebibliography}{10}

\bibitem{Tanimura_1992}
S.~Tanimura, ``{Relativistic generalization and extension to the non-Abelian
  gauge theory of Feynman’s proof of the Maxwell equations},'' {\em Annals of
  Physics}, vol.~220, p.~229–247, Dec 1992.

\bibitem{Jackiw:1984rd}
R.~Jackiw, ``{3 - Cocycle in Mathematics and Physics},'' {\em Phys. Rev.
  Lett.}, vol.~54, pp.~159--162, 1985.

\bibitem{Dai:2008bh}
P.~Dai, Y.-t. Huang, and W.~Siegel, ``{Worldgraph Approach to Yang-Mills
  Amplitudes from N=2 Spinning Particle},'' {\em JHEP}, vol.~10, p.~027, 2008.

\bibitem{1410.3346}
M.~Grützmann, J.-P. Michel, and P.~Xu, ``Weyl quantization of degree 2
  symplectic graded manifolds,'' arXiv:1410.3346, 2014.

\bibitem{Moser1965}
J.~Moser, ``On the volume elements on a manifold,'' {\em Transactions of the
  American Mathematical Society}, vol.~120, pp.~286--286, Feb. 1965.

\bibitem{Gualtieri:2007ng}
M.~Gualtieri, ``{Generalized complex geometry},'' arXiv:math/0703298, 2007.

\bibitem{Alekseev:2004np}
A.~Alekseev and T.~Strobl, ``{Current algebras and differential geometry},''
  {\em JHEP}, vol.~03, p.~035, 2005.

\bibitem{Severa:2017oew}
P.~\v{S}evera, ``{Letters to Alan Weinstein about Courant algebroids},'' 7
  arXiv:1707.00265, 2017.

\bibitem{Buscher:1987sk}
T.~H. Buscher, ``{A Symmetry of the String Background Field Equations},'' {\em
  Phys. Lett.}, vol.~B194, pp.~59--62, 1987.

\bibitem{Shelton:2005cf}
J.~Shelton, W.~Taylor, and B.~Wecht, ``{Nongeometric flux compactifications},''
  {\em JHEP}, vol.~10, p.~085, 2005.

\bibitem{Hull:2019iuy}
C.~Hull and R.~J. Szabo, ``{Noncommutative gauge theories on D-branes in
  non-geometric backgrounds},'' {\em JHEP}, vol.~09, p.~051, 2019.

\bibitem{Halmagyi:2009te}
N.~Halmagyi, ``{Non-geometric Backgrounds and the First Order String Sigma
  Model},'' arXiv:0906.2891, 2009.

\bibitem{Bouwknegt:2003vb}
P.~Bouwknegt, J.~Evslin, and V.~Mathai, ``{T duality: Topology change from H
  flux},'' {\em Commun. Math. Phys.}, vol.~249, pp.~383--415, 2004.

\bibitem{Hohm:2010pp}
O.~Hohm, C.~Hull, and B.~Zwiebach, ``{Generalized metric formulation of double
  field theory},'' {\em JHEP}, vol.~08, p.~008, 2010.

\bibitem{Andriot:2012an}
D.~Andriot, O.~Hohm, M.~Larfors, D.~Lust, and P.~Patalong, ``{Non-Geometric
  Fluxes in Supergravity and Double Field Theory},'' {\em Fortsch. Phys.},
  vol.~60, pp.~1150--1186, 2012.

\bibitem{Andriot:2013xca}
D.~Andriot and A.~Betz, ``{$\beta$-supergravity: a ten-dimensional theory with
  non-geometric fluxes, and its geometric framework},'' {\em JHEP}, vol.~12,
  p.~083, 2013.

\bibitem{Blumenhagen:2013aia}
R.~Blumenhagen, A.~Deser, E.~Plauschinn, F.~Rennecke, and C.~Schmid, ``{The
  Intriguing Structure of Non-geometric Frames in String Theory},'' {\em
  Fortsch. Phys.}, vol.~61, pp.~893--925, 2013.

\bibitem{Grana:2008yw}
M.~Grana, R.~Minasian, M.~Petrini, and D.~Waldram, ``{T-duality, Generalized
  Geometry and Non-Geometric Backgrounds},'' {\em JHEP}, vol.~04, p.~075, 2009.

\bibitem{Plauschinn:2018wbo}
E.~Plauschinn, ``{Non-geometric backgrounds in string theory},'' {\em Phys.
  Rept.}, vol.~798, pp.~1--122, 2019.

\bibitem{Heller:2016abk}
M.~A. Heller, N.~Ikeda, and S.~Watamura, ``{Unified picture of non-geometric
  fluxes and T-duality in double field theory via graded symplectic
  manifolds},'' {\em JHEP}, vol.~02, p.~078, 2017.

\bibitem{Blumenhagen:2012nk}
R.~Blumenhagen, A.~Deser, E.~Plauschinn, and F.~Rennecke, ``{A bi-invariant
  Einstein-Hilbert action for the non-geometric string},'' {\em Phys. Lett.},
  vol.~B720, pp.~215--218, 2013.

\bibitem{Jurco:2015bfs}
B.~Jur\v{c}o and J.~Vysok\'y, ``{Heterotic reduction of Courant algebroid
  connections and Einstein\textendash{}Hilbert actions},'' {\em Nucl. Phys. B},
  vol.~909, pp.~86--121, 2016.

\bibitem{Jurco:2017gii}
B.~Jurco and J.~Vysoky, ``{Poisson\textendash{}Lie T-duality of string
  effective actions: A new approach to the dilaton puzzle},'' {\em J. Geom.
  Phys.}, vol.~130, pp.~1--26, 2018.

\bibitem{Haag:1974qh}
R.~Haag, J.~T. Lopuszanski, and M.~Sohnius, ``{All Possible Generators of
  Supersymmetries of the s Matrix},'' {\em Nucl. Phys. B}, vol.~88, p.~257,
  1975.

\bibitem{Coleman:1967ad}
S.~R. Coleman and J.~Mandula, ``{All Possible Symmetries of the S Matrix},''
  {\em Phys. Rev.}, vol.~159, pp.~1251--1256, 1967.

\bibitem{Deser:2019pue}
A.~Deser, ``{Pre‐NQ Manifolds and Correspondence Spaces: the Nilmanifold
  Example},'' {\em Fortsch. Phys.}, vol.~67, no.~8-9, p.~1910006, 2019.

\bibitem{Blumenhagen:2012pc}
R.~Blumenhagen, A.~Deser, E.~Plauschinn, and F.~Rennecke, ``{Bianchi Identities
  for Non-Geometric Fluxes - From Quasi-Poisson Structures to Courant
  Algebroids},'' {\em Fortsch. Phys.}, vol.~60, pp.~1217--1228, 2012.

\bibitem{Roytenberg:2002nu}
D.~Roytenberg, ``{On the structure of graded symplectic supermanifolds and
  Courant algebroids},'' in {\em {Workshop on Quantization, Deformations, and
  New Homological and Categorical Methods in Mathematical Physics Manchester,
  England, July 7-13, 2001}}, 2002.

\bibitem{Seiberg:1999vs}
N.~Seiberg and E.~Witten, ``{String theory and noncommutative geometry},'' {\em
  JHEP}, vol.~09, p.~032, 1999.

\bibitem{Boffo:2020vqx}
E.~Boffo and P.~Schupp, ``{Dual gravity with $R$ flux from graded Poisson
  algebra},'' in {\em {19th Hellenic School and Workshops on Elementary
  Particle Physics and Gravity}}, 3 2020.

\bibitem{Boffo:2019zus}
E.~Boffo and P.~Schupp, ``{Deformed graded Poisson structures, Generalized
  Geometry and Supergravity},'' {\em JHEP}, vol.~01, p.~007, 2020.

\bibitem{lean2012dorfman}
M.~J. Lean, ``Dorfman connections and courant algebroids,'' arXiv:1209.6077,
  2012.

\bibitem{Jurco:2015ywk}
B.~Jurco, F.~S. Khoo, P.~Schupp, and J.~Vysoky, ``{Generalized geometry and
  non-symmetric gravity},'' in {\em {Proceedings, 14th Marcel Grossmann Meeting
  on Recent Developments in Theoretical and Experimental General Relativity,
  Astrophysics, and Relativistic Field Theories (MG14) (In 4 Volumes): Rome,
  Italy, July 12-18, 2015}}, vol.~3, pp.~2683--2688, 2017.

\bibitem{Asakawa:2015jza}
T.~Asakawa, H.~Muraki, and S.~Watamura, ``{Gravity theory on Poisson manifold
  with $R$-flux},'' {\em Fortsch. Phys.}, vol.~63, pp.~683--704, 2015.

\bibitem{Andriot:2012wx}
D.~Andriot, O.~Hohm, M.~Larfors, D.~Lust, and P.~Patalong, ``{A geometric
  action for non-geometric fluxes},'' {\em Phys. Rev. Lett.}, vol.~108,
  p.~261602, 2012.

\bibitem{Coimbra:2011nw}
A.~Coimbra, C.~Strickland-Constable, and D.~Waldram, ``{Supergravity as
  Generalised Geometry I: Type II Theories},'' {\em JHEP}, vol.~11, p.~091,
  2011.

\bibitem{Heller:2017mwz}
M.~A. Heller, N.~Ikeda, and S.~Watamura, ``{Courant algebroids from double
  field theory in supergeometry},'' in {\em {Proceedings, Workshop on Strings,
  Membranes and Topological Field Theory}}, pp.~315--335, WSPC, WSPC, 2017.

\bibitem{Severa:2018pag}
P.~\v{S}evera and F.~Valach, ``{Courant algebroids, Poisson-Lie T-duality, and
  type II supergravities},'' {\em Commun. Math. Phys.}, vol.~375, no.~1,
  pp.~307--344, 2020.

\end{thebibliography}

\end{document}